\documentclass[12pt,preprint]{emulateapj}
\begin{document}

\parindent=1.0cm

\title{Seeing Red in NGC 1978, NGC 55, and NGC 3109 \altaffilmark{1} \altaffilmark{2}}

\author{T. J. Davidge}

\affil{Dominion Astrophysical Observatory,
\\National Research Council of Canada, 5071 West Saanich Road,
\\Victoria, BC Canada V9E 2E7\\tim.davidge@nrc.ca}

\altaffiltext{1}{Based on observations obtained at the Gemini Observatory, which is
operated by the Association of Universities for Research in Astronomy, Inc., under a
cooperative agreement with the NSF on behalf of the Gemini partnership: the National
Science Foundation (United States), the National Research Council (Canada), CONICYT
(Chile), Minist\'{e}rio da Ci\^{e}ncia,
Tecnologia e Inova\c{c}\~{a}o (Brazil) and Ministerio de Ciencia, Tecnolog\'{i}a e
Innovaci\'{o}n Productiva (Argentina).}

\altaffiltext{2}{This research has made use of the NASA/IPAC Infrared Science Archive,
which is operated by the Jet Propulsion Laboratory, California Institute of Technology,
under contract with the National Aeronautics and Space Administration.}

\begin{abstract}

	Spectra of the intermediate age star cluster NGC 1978 and the dwarf 
irregular galaxies NGC 55 and NGC 3109 are discussed. The spectra were 
recorded with the Gemini Multi-Object Spectrograph on Gemini South and 
span the 0.7 to $1.1\mu$m wavelength interval. Five slit pointings were observed in 
NGC 1978, and these are used to examine stochastic effects on the integrated red light 
from an intermediate age cluster. The removal of either the brightest M giant or 
the brightest C star from the co-added spectrum has minor affects on the equivalent 
withs of the Ca triplet. The most robust signature of C stars in the integrated 
cluster spectrum at these wavelengths is the CN band head near 7900\AA. The equivalent 
widths of Ca triplet lines in the NGC 1978 spectrum and in the spectra of 
individual cluster stars are larger than expected for a scaled-solar 
abundance system. It is suggested that these stars have a lower than expected surface 
gravity, which might occur if the stars in NGC 1978 have been subject to extra 
mixing processes, as suggested by Lederer et al. (2009, A\&A, 502, 93). 
The near-infrared color profile of NGC 1978 is shown to contain a prominent red cusp 
in the central 10 arcsec, and the suppression of light from this cusp does not affect 
the depth of the Ca lines in the integrated spectrum. The NGC 55 spectra run 
parallel to the major axis, and a gradient is found in 
the strength of the Ca lines, in the sense that the Ca lines 
weaken with increasing distance from the disk plane. Comparisons with 
models suggest that the disk light is dominated by stars with ages $1 - 2$ Gyr, in 
agreement with star-forming histories (SFHs) obtained from the analysis of CMDs. 
The NGC 55 spectra also sample a large star-forming complex.
The age of this complex inferred from comparisons with models is broadly 
consistent with that estimated from a near-infrared CMD of the same region. 
The CN band head at 7900\AA\ in this part of NGC 55 is detected, but this a 
signature of red supergiants (RSGs) rather than C stars.  
The NGC 3109 observations sample three different parts of 
that galaxy but have a low signal-to-noise ratio. Comparisons 
with models suggest that the light from the NGC 3109 disk 
at red wavelengths is dominated by RSGs with ages of at most a few tens 
of Myr, in qualitative agreement with SFHs that are based on photometric measurements. 

\end{abstract}

\keywords{galaxies: individual (NGC 55 and NGC 3109) -- globular clusters: individual 
(NGC 1978) -- galaxies: stellar content}

\section{INTRODUCTION}

	The stellar contents of distant galaxies and the crowded central regions 
of the nearest galaxies can only be investigated through studies of their 
integrated light. Spectroscopic studies of integrated light at rest frame blue 
and visible wavelengths are of prime importance for such efforts, as there are 
well-calibrated features at these wavelengths that probe chemical content and 
age (e.g. Worthey 1994). However, information that is important for understanding 
stellar content is also present at other wavelengths. 

	Highly evolved stars in intermediate 
age and old systems emit much of their light at red/near-infrared (NIR) 
wavelengths, allowing these objects to be observed efficiently at wavelengths 
(1) that are less prone to extinction than objects where the light output peaks at 
shorter wavelengths, and (2) where corrections for distortions to 
the incident wavefronts can produce a significant improvement in image quality.
Studies of integrated light at red/NIR wavelengths also facilitates 
searches for spectroscopic signatures of C stars. 
The detection of C stars in integrated light is of interest as these objects 
probe epochs that bridge the most recent and earliest episodes of star formation. 
Model spectra generated by Maraston (1998; 2005) and for the Bag of Stellar Tricks and 
Isochrones (Cordier et al. 2007) show that prominent spectroscopic signatures 
of C stars might be expected in the spectra of stellar systems that span a range of 
ages and metallicities. CN bands from C stars have been detected in the NIR spectrum 
of the star cluster W3 in NGC 7252 (Mouhcine et al. 2002).
The Ballick-Ramsey C$_2$ band head near $1.76\mu$m has been detected in 
the NIR spectrum of the dwarf lenticular galaxy NGC 5102 (Miner et al. 2011) and 
the nearby elliptical galaxy Maffei 1 (Davidge et al. 2015). A tentative detection 
of the C$_2$ band head near 1.76$\mu$m has also been made in the central regions of the 
starburst galaxy NGC 253 by Davidge (2016), hinting at elevated levels of star-forming 
activity near the center of that galaxy for at least a few hundred Myr. 

	Are the stellar contents inferred from the 
analysis of integrated light consistent with those found from 
the analysis of CMDs? By necessity, answering this question involves 
studies that are restricted to nearby star clusters and galaxies, typically focusing 
on systems with moderately high surface brightnesses (e.g. Davidge \& Jensen 2007). 
In the present study, long slit spectra recorded with the Gemini 
Multi-Object Spectrograph (GMOS) on Gemini South are used to investigate the 
red spectroscopic characteristics of the intermediate age star 
cluster NGC 1978 in the LMC and the dwarf irregular galaxies NGC 55 and NGC 3109. 
While the comparatively low surface brightnesses of NGC 55 and NGC 3109 present 
observational challenges for studying their integrated light, the surface brightnesses 
are such that stars with a range of intrinsic brightnesses can be resolved 
to construct CMDs.

	The GMOS spectra cover the wavelength interval $0.7 - 1.1\mu$m. 
The spectra thus sample wavelengths where the molecular 
bands that are spectroscopic signatures of cool, highly evolved stars are evident. 
The lines of the NIR Ca triplet, which are prominent 
features in the spectra of late-type stars and are sensitive to surface gravity 
and metallicity (e.g. Jones et al. 1984; Armandroff \& Zinn 1988, and numerous 
subsequent studies) also fall in this wavelength range, as do numerous transitions of 
the Paschen series of Hydrogen.

	NGC 1978 is an intermediate age cluster in the LMC that 
was selected for study because of its metallicity and age. 
With [Fe/H] $\sim -0.4$ (Ferraro et al. 2006) NGC 1978 has a 
metallicity that is comparable to that found in young stars in 
NGC 55 (see below). Mucciarelli et al. (2007) determine 
an age of $1.9 \pm 0.1$ Gyr from a deep CMD, making NGC 1978 one of the oldest 
of the compact intermediate age clusters in the LMC. 
The CMD of NGC 1978 does not show the extended MSTO that is 
seen in the CMDs of many other intermediate age LMC clusters (Milone et al. 2009). 
That NGC 1978 has a clearly defined MSTO -- coupled with a tiny upper limit to 
any star-to-star dispersion in metallicity (Ferraro et al. 2006) -- suggests that it 
is an unambiguous example of an intermediate age simple stellar population (SSP).

	Stellar systems with the same age and metallicity as NGC 1978 are expected to 
produce C stars (e.g. Marston 2005), although the number that might be present at any 
one time is subject to stochastic effects given the rapid pace of C star evolution. 
NGC 1978 is sufficiently massive to contain a rich population of bright 
red giants, some of which are C stars (e.g. Lloyd Evans 1980a). 
Frogel et al. (1990) discuss NIR photometry of the brightest 
stars in and around the cluster, including 9 confirmed or suspected C stars. 
Lederer et al. (2009) find that the abundance patterns in six of the C stars are not 
consistent with stellar evolution models. 

	NGC 55 is morphologically similar to the LMC and -- with NGC 
300 - belongs to one of the nearest ensembles of galaxies outside of the Local Group 
(Karachentsev et al. 2003). Castro et al. (2012), Kudritzki et al. (2016), and Patrick 
et al. (2017) examine the metallicity of the brightest blue and red supergiants 
(RSGs) in NGC 55 and find a mean metallicity that is comparable to that of the LMC.
The SFH of NGC 55 has not yet been charted in detail, although broad conclusions can be 
drawn from extant data. The deep CMDs of multiple fields in NGC 55 presented 
by Dalcanton et al. (2009) reveal a bright main sequence, red and blue core 
helium-burning sequences and red giant branches. Together, these indicate that star 
formation has occured over a broad range of epochs. Davidge (1998; 2005) and 
Seth et al. (2005a) conclude that there has been significant star formation in NGC 55 
during the past 0.1 -- 0.2 Gyr. Narrow-band images examined by Graham \& 
Lawrie (1982) and Ferguson et al. (1996) show a complex ISM that 
contains numerous bubbles and structures that are indicative of 
vigorous recent star-forming activity, much of which at present is concentrated in two 
large star-forming complexes. The structural complexities of the NGC 55 ISM are 
consistent with the large rates of mass accretion and outflow deduced 
for the galaxy by Kudritzki et al. (2016) from chemical evolution models. 

	While there is evidence of recent star formation, 
Davidge (2005) discusses evidence that the 
present-day SFR of NGC 55 is lower than that at earlier epochs. 
Of particular note is the large population of luminous AGB stars. 
NGC 55 was one of the first galaxies outside of the Local Group to be surveyed for 
C stars (Pritchet et al. 1987). The C star frequency is roughly 
an order of magnitude higher than in the LMC or SMC (e.g. Figure 4 of Battinelli 
\& Demers 2005). Given that NGC 55 and the LMC have similar metallicities, 
then -- if it assumed that there was continuous star-forming activity during 
intermediate epochs -- this argues that NGC 55 had a much higher SFR during 
intermediate epochs than the LMC. 

	NGC 3109 is a Magellanic irregular galaxy that is the dominant 
member of a structurally ordered (Bellazzini et al. 2013) mini-group that may or 
may not belong to the Local Group. Other candidate members of this group are 
the dwarf galaxies Antlia A, Sextans A, Sextans B, and Leo P. A recent addition 
to the group is the dwarf galaxy Antlia B (Sand et al. 2015). 

	Hosek et al. (2014) discuss spectra of some of the brightest stars in NGC 
3109, and find [Fe/H] $\sim -0.7$ dex; NGC 3109 is thus $\sim 0.3$ dex more metal-poor 
than either NGC 55 or NGC 1978. Mateo (1998) presents a schematic SFH based on 
early photometric studies that suggests a high SFR at early epochs, with reduced -- but 
non-negligible -- star-forming activity during subsequent epochs. 
A rich population of C stars is present (Demers et al. 2003), indicating that 
there was significant star-forming activity during 
intermediate epochs. The C star frequency in NGC 3109 is 
similar to that in NGC 55 (Battinelli \& Demers 2005). As NGC 55 has a 
higher metallicity than NGC 3109 then NGC 55 would be expected to have a {\it lower} C 
star frequency if the two galaxies had identical SFHs during intermediate epochs. 
The similarity in C star frequencies thus suggests that (1) the SFHs of these 
galaxies during intermediate epochs were different, and (2) the time-averaged 
SFR during epochs when C stars are produced was lower in NGC 3109 than in NGC 55.

	The goal of the present study is to examine the spectroscopic properties of 
the integrated light from all three systems. The spectra are compared with models 
from the EMILES compilation described by Rock et al. (2016), with the goal of 
finding luminosity-weighted ages, which in turn are compared with 
expectations from CMDs. The properties of individual stars 
and star clusters that are sampled with these data are also investigated. 

	The paper is structured as follows. A description of the observations and the 
procedures used to reduce the data are presented in Section 2. 
Stochastic variations in stellar content at red wavelengths and the influence 
of C stars on integrated spectra are examined with the NGC 1978 data 
in Section 3. The spectra of NGC 55 and NGC 3109 are the subjects of Sections 4 and 5. 
The SFHs of NGC 55 and NGC 3109 suggest that the 
integrated spectra of these galaxies should show marked differences, and these are seen.
A discussion and summary of the results follows in Sections 6 and 7.

\section{OBSERVATIONS \& REDUCTIONS}

\subsection{Description of Observations: NGC 1978 and NGC 3109}

	Spectra of NGC 1978 and NGC 3109 were recorded 
with GMOS (Hook et al. 2004) on Gemini South on four nights in 2013 
and 2014 as part of program GS-2012B-Q-93 (PI: Davidge). 
The detector in GMOS at that time was a mosaic of three EEV CCDs, 
each of which had a $6144 \times 4068$ pixel format, with $13.5\mu$m square 
pixels. Each pixel subtended 0.073 arcsec on a side. 

	Light was dispersed with the R400 ($\lambda_{blaze} = 0.76\mu$m, 
400 lines/mm) grating, with an RG610 filter deployed to 
suppress light from higher orders. The grating was rotated so that the 
wavelength at the center of the detector array was $0.9\mu$m. The slit width 
was set at 2 arcsec, and the spectral resolution of sources that uniformly 
fill the slit is then $\sim 500$ at the blaze wavelength. 

	The spectra were recorded with $4 \times 4$ pixel binning during read-out. 
As the program was intended for poor ($> 1$ arcsec) 
seeing conditions then a factor of four binning in the spatial direction 
provides reasonable sampling of point sources. 
This binning also does not degrade the spectral resolution. 

	A rectangular-shaped region that runs through the 
center of NGC 1978 was mapped with five overlapping slit placements. The GMOS slit 
runs parallel to the major axis of this highly flattened 
(eccentricity = 0.3; Fischer et al. 1992, Mucciarelli et al. 2007) cluster. 
The offset between each slit position is 1.5 arcsec along the minor axis, and so 
an 8 arcsec wide region that parallels the major axis is mapped. 

	Spectra that sample three different locations were recorded of NGC 3109, 
with the slit more-or-less parallel to the minor axis. 
The slit was stepped in 1 arcsec increments between exposures. 
A 4 arcsec wide band is covered for two of the 
pointings (P1 and P2), while time limitations meant that only a 3 arcsec wide region 
was mapped for the third pointing (P3).

	The slit placements for NGC 3019 were selected to sample structures 
near the major axis of the galaxy that were flagged as 
candidate compact intermediate age clusters based on their appearance in 
images at ultraviolet and mid-infrared wavelengths. 
Efforts to detect these objects in the current program had 
poor success. Therefore, the discussion of NGC 3109 is limited to the integrated 
spectrum and resolved stars. 

	The approximate mid-points of the slit placements for NGC 1978 and NGC 3109 
are indicated in Figure 1. The images in Figure 1 are in the $r'$ 
filter and were taken with GMOS for target acquisition. 
Additional details of the observations, including 
the co-ordinates of the field centers, are listed in Table 1. 

\begin{figure}
\figurenum{1}
\epsscale{1.0}
\plotone{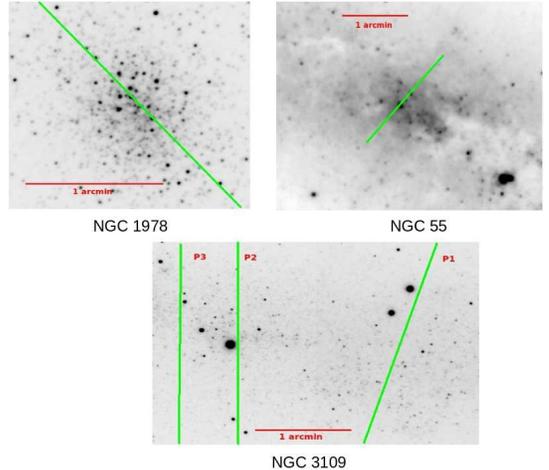}
\caption{$r'$ images of NGC 1978, NGC 55, and NGC 3109 that were recorded 
with GMOS immediately before spectra were obtained. North is at the top, and East 
is to the right. The initial location and orientation of the slit for 
each object is shown as a green line. The slit was 
offset perpendicular to the long axis in 1.5 
arcsec increments for NGC 1978 and 1 arcsec increments for NGC 3109 
to map out wider regions than would otherwise be covered by a single 
pointing. The effective slit length for NGC 55 is 
shorter than for the other targets because the galaxy was 
observed at two locations on the slit (see text), and only the area of common 
overlap is shown.}
\end{figure}

\begin{deluxetable}{lcccc}
\tablecaption{Details of Observations}
\startdata
\tableline\tableline
Target & RA\tablenotemark{a} & Dec\tablenotemark{a} & Date & Exposure \\
 & (2000) & (2000) & (UT) & (sec) \\
\tableline
NGC 1978 & 05:28:45.2 & --66:14:11.9 & 2013--01--25 & $5 \times 300$ \\
 & & & & \\
NGC 3109 (P1) & 10:03:10.9 & --26:09:42.3 & 2014--01--28 & $3 \times 900$ \\
NGC 3109 (P2) & 10:03:04.4 & --26:09:42.3 & 2013--03--11 & $2 \times 900$ \\
 & & & 2014--04--20 & $1 \times 900$ \\
NGC 3109 (P3) & 10:03:02.1 & --26:09:39.7 & 2014--01--28 & $2 \times 900$ \\
 & & & & \\
NGC 55 & 00:14:53.6 & --39:11:47.9 & 2017--07-03 & $4 \times 450$ \\
\tableline
\enddata
\tablenotetext{a}{Slit center.}
\end{deluxetable}

	Calibration frames were recorded 
with the same instrumental configuration as the science observations. 
Off-sky calibrations consisted of bias frames, CuAr arcs, 
and images of dispersed light from a flat-field light source. 
The arc and flat-field light sources are located in the 
Gemini Facility Calibration Unit (`GCAL'). Following standard 
Gemini observing protocol, the first two were recorded 
during daylight hours immediately after the nights when science data were recorded. 
The flat field frames were interlaced with the on-sky observations. 
On-sky calibrations consisted of spectra of early-type stars that were observed 
to monitor telluric absorption features and system throughput.

\subsection{Description of Observations: NGC 55}

	Spectra of NGC 55 were recorded with GMOS on Gemini South 
during the night of July 4, 2017 as part of program GS-2017A-Q-98 
(PI: Davidge). The detector was a mosaic of three 
Hamamatsu CCDs, which replaced the EEV CCDs that were used for the NGC 1978 and 
NGC 3109 observations. Each Hamamatsu CCD has a $2048 \times 4166$ pixel format, 
with $15\mu$m square pixels. 

	Hardy et al. (2012) describe the characteristics of the Hamamatsu CCDs.
These devices have high quantum efficiencies over a broad wavelength range, with
quantum efficiencies exceeding 80\% between 0.55 and $0.9\mu$m. 
The Hamamatsu CCDs deliver almost an order of magnitude gain in quantum efficiency at 
$1\mu$m when compared with their EEV predecessors ($\sim 40\%$ for the 
Hamamatsu CCDs versus $\sim 5\%$ for the EEVs) \footnote[1]
{http://www.gemini.edu/sciops/instruments/gmos/imaging/detector-array/gmosn-array-hamamatsu}.

	The slit was oriented to parallel the minor axis of NGC 55 and pass 
through the area observed by Davidge (2005) in $J, H$ and $K$. The location of the 
slit is shown in Figure 1. Details of the NGC 55 observations are 
summarized in Table 1. 

	The use of a different detector aside, the configuration of GMOS 
for the NGC 55 observations followed that described in Section 2.1. 
The spectra were recorded with $4 \times 4$ pixel 
binning during read-out, with the light dispersed by the 
R400 grating. A 2 arcsec wide slit was used, with an 
OG515 filter deployed to suppress signal from higher orders. 
A series of calibration exposures that parallel those obtained for NGC 1978 
and NGC 3109 were also recorded. 

	There are three departures from the observing procedures described in 
Section 2.1. First, the NGC 55 spectra were recorded in pairs having central 
wavelengths of 0.82 and $0.85\mu$m. Spectra recorded with two central wavelengths 
make it possible to fill the gaps between the detector arrays during processing. 
Second, the telescope pointing was not offset perpendicular to the slit
between exposures, and so only a 2 arcsec wide band through NGC 55 is sampled. 
Finally, the NGC 55 spectra were recorded with the major axis of the galaxy offset 
between two locations along the slit to facilitate the suppression of sky lines. 
While this has proven to be an effective means of suppressing strong telluric 
emission lines in the NIR, the final spectra do not sample the entire 5.5 arcmin 
length of the GMOS slit, but instead are restricted to $\sim 100$ arcsec. 

\subsection{Reduction of Observations and Extraction of Spectra}

	The data were processed using standard reduction 
procedures for long-slit spectra. To begin, the output 
from all CCDs was multiplied by their gain values. 
A master bias frame for each day was constructed by median-combining individual 
bias frames. These were subtracted from the science exposures, after which 
overscan regions were trimmed from all frames. Cosmetic defects were 
suppressed by interpolating between pixels. Cosmic rays were identified using 
an edge detection algorithm, with interpolation used to fill affected areas.

	The next step was to remove pixel-to-pixel differences in sensitivity and 
correct for non-uniform throughput along the slit (`flat-fielding'). 
The flat-field frames recorded for each set of observations 
were processed using the steps described in the preceeding paragraph 
and the results were scaled to produce a mean signal of unity. 
The flat-field correction was applied by dividing the 
bias-subtracted and cosmetically-cleaned science data by the processed flats. 

	Geometric distortions were removed from the flat-fielded data 
using a reference grid that was defined by tracking emission lines in the CuAr spectra. 
The sky was then removed from the distortion-corrected images. 
For NGC 1978 and NGC 3109 the mean background level was measured 
near the ends of the slit, and this was subtracted from the two-dimensional 
spectra on a row-by-row basis. As NGC 55 was observed 
with the galaxy positioned at two different locations on 
the slit, background subtraction was achieved simply by subtracting successive 
observations in which the galaxy was offset along the slit. These 
difference frames contain positive and negative background-subtracted versions 
of the NGC 55 spectrum. The spatial intervals along the slit where the two spectra 
did not overlap were extracted. The negative spectrum was 
aligned with the positive version, and subtracted from the positive 
spectrum to produce a summed spectrum. There are two such summed pairs for NGC 55, 
and these were averaged together.

	Spectra of the main bodies of NGC 1978 and NGC 3109 were obtained by combining 
the signal within the FWHM of light profiles that were constructed by collapsing 
the signal along the dispersion axis between 0.7 and 0.9$\mu$m. While the S/N ratio 
of the NGC 3109 spectra are too low to allow spectra that sample different 
spatial intervals from the major axis to be extracted, 
this was not the case for NGC 55. Spectra of NGC 55 were extracted 
by combining signal within three angular intervals 
from the major axis, using the light profile obtained 
by co-adding signal between 0.7 and $0.9\mu$m as a guide.

	Spectra of individual stars were extracted 
from the NGC 1978 and NGC 3109 observations -- this was not possible 
for NGC 55 given the distance to this galaxy, 
which complicates efforts to detect individual stars. Some stars 
in the NGC 1978 and NGC 3109 spectra are obvious blends, 
and spectra were only extracted when contaminating objects were 
at least a factor of two fainter than the dominant star. 
The extracted spectra contain contributions from both the star and 
unresolved light from the main body of the host. The latter was removed by 
averaging signal on either side of each target to construct a local sky spectrum 
and then subtracting the result from the stellar spectrum.

	Extracted spectra were wavelength calibrated 
using a dispersion solution obtained from the CuAr arcs. 
The wavelength response of the spectra at this stage of the processing 
contains contributions from the flat-field 
lamp, the instrument and telescope optics, the atmosphere, 
and the source itself. The first three were removed by dividing 
the extracted spectra by spectra of the telluric standard stars. The 
telluric star spectra used in this step had been processed using the sequence 
described above, and then scaled to a mean flux of unity. 
The telluric star observed for the NGC 1978 and NGC 
3109 observations -- LTT 3864 -- has an F spectral-type. Weak absorption lines 
from the Ca triplet are present in its spectrum and these were removed by fitting Vogt 
functions to the appropriate wavelength intervals and subtracting the fitted 
profiles from the original spectrum. The O subdwarf Fiege 110 served as 
the telluric reference star for NGC 55, and absorption features are negigible in the 
spectrum of that star.

	Division by the spectrum of a telluric standard simplifies the wavelength 
response, leaving a response that is normalized to the spectral-energy 
distribution (SED) of the telluric star. Given that the telluric standards have a 
spectrum that is free of strong absorption features then this means that 
only smooth, gradual changes in response with 
wavelength are present. To remove this residual response, 
each spectrum was divided by a low-order continuum function that was fit to that 
spectrum. Restricting the fitting function to a low order reduces the possibility 
that molecular absorption bands and deep atomic features will skew the continuum 
location. The final step was to shift the continuum-corrected spectra 
into the rest frame. It is these continuum and 
Doppler-corrected spectra that are discussed in this paper.

\section{RESULTS: NGC 1978}

\subsection{Stellar Spectra}

	A large fraction of the red and NIR light from intermediate age 
star clusters originates from highly evolved stars, and 
so the properties of these objects play a significant role in defining the SEDs 
of their host systems. Therefore, the spectra of stars that were extracted from 
the NGC 1978 observations are examined before considering the integrated spectrum 
of NGC 1978. While the stellar spectra have a low wavelength resolution and only 
limited information can be gleaned from them individually, when considered in concert 
they are valueable aids for interpreting the light from NGC 1978. Emphasis is placed 
on two aspects of these stars: (1) preliminary spectral-types, based largely 
on the depths of molecular features and NIR colors, and (2) the depths of 
the Ca triplet absorption lines.

\subsubsection{Spectral Types}

	The high stellar density in NGC 1978 limits the sample 
of objects for which spectra can be extracted to the brightest cluster members and 
bright field stars (i.e. members of the LMC that do not belong to NGC 1978, as 
well as foreground Galactic stars). The spectra of the brightest stars in the area 
sampled by GMOS are shown in Figure 2, where the stars are ordered 
according to the depth of TiO and CN absorption near $0.71\mu$m.
With one exception -- the most luminous star in the cluster (bottom spectrum 
in Figure 2) -- the stars in Figure 2 have brightnesses that are within a 
factor of 2 of each other near $0.8\mu$m, as gauged from the signal in the spectra near 
this wavelength. 

	All of these stars are in the 2MASS Point Source 
Catalog (Skrutskie et al. 2006), and the 2MASS identifications and NIR brightnesses 
and colors are listed in the top part of Table 2. As in Figure 2, these stars 
are ordered according to the depth of TiO/CN absorption near $0.71\mu$m.
Highly evolved stars can be photometric variables, and the photometry in 
Table 2 is based on only a single epoch. Still, it is apparent that the 
ordering based on depth of the TiO and CN absorption near $0.71\mu$m correlates roughly 
with $K$ brightness, in the sense that the faintest stars have the weakest 
band heads.

	Radial velocities are listed in the last column of Table 2. These were 
measured by cross-correlating with stars from the Cenarro et al. (2001) database. 
These velocities should be considered as preliminary only, due to the 
relatively low spectral resolution of the GMOS data and the potential 
for errors arising from non-uniform illumination of the slit (`guiding errors'). 
In addition, radial velocity standards were not observed as part of this program. 
The estimated uncertainties in the velocities are $\pm 30$ km sec$^{-1}$. A radial 
velocity was not measured for 2MASS05284449--6614039 as there was not a suitable 
template for cross correlation. Excluding 2MASS05283756--6613040, the mean of the 
radial velocities in Table 2 is $280 \pm 10$ km sec$^{-1}$, where the uncertainty 
is the standard error of the mean. The mean velocity agrees 
with other measurements of the LMC velocity (e.g Richter et al. 1987) 
and the NGC 1978 velocity found by Fischer et al. (1992).

\begin{figure}
\figurenum{2}
\epsscale{1.0}
\plotone{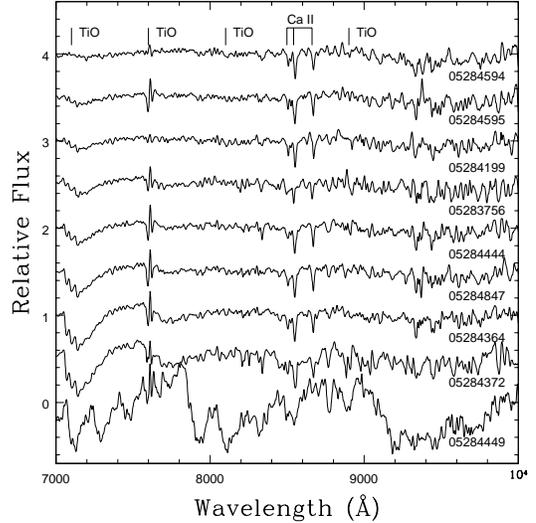}
\caption{Spectra of the brightest stars observed in NGC 1978. The stars are 
ordered according to the depth of the TiO/CN band head near $0.71\mu$m, from weakest 
to strongest. Each star is labelled with the first half of its 2MASS Point 
Source Catalog identifier -- the complete names are listed in Table 2. 
The $J-K$ colors of many of these stars are consistent with an 
early M giant spectral-type. 2MASS05284847--6614387 is star \# 8 in the numbering 
scheme of Lloyd Evans (1980b), and was one of two stars observed by Olszewski et al. 
(1991) to estimate the metallicity of NGC 1978 from the depth of Ca triplet lines. 
2MASS05283756--6613040 is a foreground Galactic star (see text), while 
2MASS05284449--6614309 is a previously identified C star.}
\end{figure}

\begin{deluxetable}{crccccc}
\tablecaption{Properties of stars in NGC 1978.}
\startdata
\tableline\tableline
2MASS \# \tablenotemark{a} & $K$ & $J-K$& $H-K$ & CaT & Spectral & Radial Velocity \tablenotemark{b}\\
 & & & & (\AA) & Type & (km sec$^{-1}$ \\
\tableline
05284594--6614138 & 12.523 & 0.888 & 0.218 & 10.37 & KIII & 340 \\ 
05284595--6614194 & 12.479 & 1.071 & 0.228 & 10.82 & KIII & 250 \\ 
05284199--6613457 & 12.498 & 1.034 & 0.159 & 10.08 & KIII & 360 \\ 
05283756--6613040 & 13.324 & 0.776 & 0.154 & 5.83 & K/M V & 20 \\ 
05284444--6613599 & 11.802 & 1.115 & 0.278 & 9.91 & MIII & 230 \\ 
05284847--6614387 & 11.916 & 1.102 & 0.246 & 9.99 & MIII & 240 \\ 
05284364--6613530 & 11.387 & 1.140 & 0.255 & 11.71 & MIII & 220 \\ 
05284372--6614037 & 11.199 & 1.148 & 0.290 & 11.00 & MIII & 290 \\ 
05284449--6614039 & 9.676 & 1.814 & 0.667 & 6.33 & C & -- \\ 
 & & & & \\
05284934--6614486 & 14.981 & 0.862 & 0.168 & 10.14 & KIII & 240 \\ 
05284668--6614262 & 14.877 & 0.571 & 0.024 & 7.54 & GIII? & 260 \\ 
05283704--6612590 & 14.368 & 1.056 & 0.265 & 10.29 & KIII & 320 \\ 
05285300--6615148 & 13.632 & 0.865 & 0.073 & 10.05 & KIII & 270 \\ 
05284317--6613564 & 13.452 & 1.001 & 0.314 & 11.08 & KIII & 280 \\ 
05284570--6614084 & 13.030 & 0.759 & 0.175 & 12.02 & KIII & 290 \\ 
\tableline
\enddata
\tablenotetext{a}{The stars in the upper part of the table belong to the bright 
sample, and are ordered according to the depth of the TiO/CN band head near $0.71\mu$m, 
from weakest to strongest. The stars in the lower part belong to the faint sample, 
and are ordered according to $K$ magnitude, from faintest to brightest.}
\tablenotetext{b}{Estimated uncertainties are $\pm 30$ km sec$^{-1}$.}
\end{deluxetable}

	Six of the spectra in Figure 2 have prominent TiO band heads 
at 7100\AA\ and deep Ca triplet lines. Stars with spectral-type M0III have $J-K = 1.01$ 
(Bessell \& Brett 1988), which corresponds to $J-K \sim 1.05$ in the 2MASS photometric 
system (Carpenter 2001). Based on the depths of the TiO band heads, coupled 
with the $J-K$ colors in Table 2, and assuming that $E(J-K) = 0.03$ for the LMC, 
then many of the stars in Figure 2 are probably early M giants.
Based on the depth of the TiO bands and $J-K$ color, 2MASS05284372--6614037 
has the latest spectral-type of the M giants.

	Not all of the stars in Figure 2 are M giants and -- based on the depth of the 
TiO bands and $J-K$ colors -- three are likely K giants. 
It is also evident that the TiO $0.71\mu$m band heads in the 
spectra of 2MASS05284199--6613457 and 2MASS05283756--6613040 have similar depths, 
but that the Ca triplet lines in the spectrum of the latter are noticeably weaker than 
in the former. 2MASS05283756--6613040 also has a bluer J--K color than 
2MASS05284199--6613457, and is $\sim 150$ arcsec from the cluster center, 
placing it well outside the main body of the cluster. Perhaps most significantly, 
it has a radial velocity that is not consistent with 
membership in the LMC. 2MASS05283756--6613040 is thus a Galactic field star. 

	2MASS05284449--6614039 is star \# 3 in the numbering scheme of 
Lloyd Evans (1980b) and star 7 in the Frogel et al. (1990) compilation of NIR 
photometry of bright stars in NGC 1978. In contrast to all of the other bright 
stars, the spectrum of this star is dominated by the CN bands that 
are prominent signatures of a C star. The CN band head near 
7900\AA\ , which is clearly seen in the 2MASS05284449-6614039 spectrum, is 
the feature that serves as the identifier for C stars in the narrow-band 
photometric systems described by Richer et al. (1984) and Cook et al. (1986). 
In Section 3.2 it is shown that this band head also serves as a marker for 
C stars in integrated light.

	Spectra were also extracted for a modest sample of stars that 
were selected to be up to $2$ magnitudes fainter than the brightest stars. 
Magnitudes were determined from the mean signal near 8000\AA, which serves as an 
approximate proxy for $I-$band magnitudes, in which stars near the RGB-tip show 
little sensitivity to metallicity. Stars in this second sample were restricted to 
those that have their peak signal in one of the three central slit 
positions so that all of their light is captured -- light from stars with peak 
signal in the two outermost pointings may fall outside of the area observed with GMOS 
depending on the placement of the star in the slit. All of the stars in this second 
sample are in the 2MASS PSC, and the 2MASS names and NIR photometry for these sources 
are listed in the bottom part of Table 2. The spectra of stars in this 
sample have a lower S/N ratio than those shown in Figure 3, complicating efforts 
to measure the depths of the TiO and CN features near $0.71\mu$m in individual 
spectra. However, there is a rough correlation between $K$ magnitude 
and the depth of the $0.71\mu$m TiO band head amongst the 
brightest stars listed in the top part of Table 2, and so the fainter 
stars listed in the lower part of Table 2 are 
ordered from faintest to brightest $K$ magnitude. 

	Based on the $J-K$ colors in Table 2, most of the stars 
in the fainter sample are expected to be K giants, and 
this is borne out by the mean spectrum of these objects, which is shown in 
Figure 3. The mean spectrum of the four giants in Figure 2 that have 
$K$ between 11 and 12, and so are the most evolved oxygen-rich objects 
observed here, is also shown. There is a weak discontinuity near the $0.71\mu$m 
TiO band head in the mean spectrum of fainter objects in Figure 3, and no 
other TiO band heads are visible. The Ca triplet lines in the mean spectrum are 
also slightly weaker than in the mean bright giant spectrum. The mean 
spectrum of faint objects is thus consistent with a spectral-type earlier 
than that of the M giants in Figure 2. 

\begin{figure}
\figurenum{3}
\epsscale{1.0}
\plotone{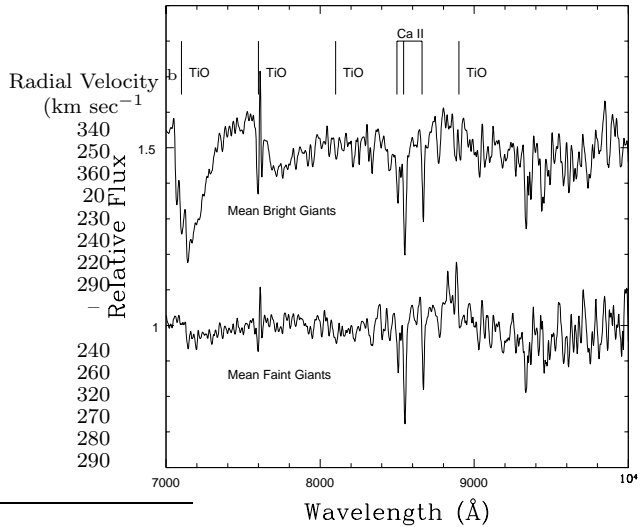}
\caption{Mean spectra of stars in NGC 1978 that have different brightnesses. The 
majority of the stars used to construct the bottom spectrum are likely early 
to mid-K giants. This conclusion is based on the weak or abscent TiO band heads 
in the mean spectrum, the strength of the Ca triplet lines, and $J-K$ colors.}
\end{figure}

\subsubsection{CaT Indices}

	The lines of the Ca triplet are among the most prominent atomic features 
in the NIR, and are probes of stellar content in integrated light. It is thus 
of interest to examine if the depths of the Ca triplet lines in the NGC 1978 stellar 
spectra are consistent with what is expected for their spectral type and 
metallicity. Cenarro and collaborators have compiled an extensive 
library of Ca triplet measurements, and the CaT indices of stars in NGC 1978 
were measured using the pass bands defined by Cenarro et al. (2001a). 

	The absolute calibration of line indices depends on factors such as spectral 
resolution, wavelength sampling, and the method used to identify the continuum. 
Cenarro et al. (2001a) discuss CaT indices measured from a large stellar library, 
and the spectra from which those indices were measured can be used to examine the 
effect of spectral resolution, wavelength sampling, and continuum 
removal on the CaT index. For the present study, two sets of 
stars in the Cenarro et al. (2001a) stellar library are considered: (1) 
K giants with [Fe/H] between --0.2 and --0.4, and (2) M giants 
with solar metallicity. The stars in the first group have spectral-types and 
metallicities that roughly match those expected for the fainter sample of giants 
in NGC 1978, as well as the stars in Figure 2 that have weaker TiO features. 
There are 6 stars in the Cenarro et al. library that match the 
criteria for the first set, and these have effective temperatures (T$_{eff}$) 
between 4000 and 4400 K, and log(g) between 1.0 and 2.3. The stars in the 
second set have spectral-types that match the stars in Figure 2 that have deep TiO 
features, although it was necessary to consider objects with solar metallicity as the 
Cenarro et al. (2001a) library does not contain metal-poor M giants. There are 
four stars in the Cenarro et al. library that match the criteria of the second set, 
and these have T$_{eff}$ between 3200 and 3800 K, and log(g) between 0.2 and 1.1.

	Spectra were downloaded from the Cenarro et al. (2001a) library 
\footnote[2]{http://www.iac.es/proyecto/miles/pages/stellar-libraries/cat-library.php}
and these were smoothed and re-sampled to match the spectral resolution and wavelength 
sampling of the GMOS spectra. As with the GMOS spectra, the continuum was 
removed by applying a low order fit. CaT indices were then measured from the 
smoothed/re-sampled spectra, and compared with the fiducial values. For K giants 
the CaT indices measured from spectra with the same resolution and sampling as 
the GMOS observations differ from the published values by $\sim 0.3$ \AA, 
in the sense that the indices obtained from the GMOS spectra are smaller. The 
difference between the two sets of measurements becomes smaller yet when M giants are 
considered, amounting to $\leq 0.1$ \AA. These comparisons suggest 
that the GMOS instrumental CaT indices are {\it more-or-less} in the 
standard system to within a few tenths of an \AA , at least when considering 
K giants or early M giants. The modest sensitivity of the CaT index to 
spectral resolution found here is consistent with the velocity 
broadening experiments that are discussed in Section 4.4.2 of Cenarro et al. (2001a).

	Instrumental CaT indices for all extracted stars are listed in Table 2. 
CaT indices in the system described by Cenarro et al. (2001a) can be found by adding 
0.1\AA\ to the instrumental indices of 2MASS05284372--6614037, 2MASS05284444--6613599, 
2MASS05284364--6613530, and 2MASS05284847--6614387, as these are likely 
early M giants. Similarly, 0.3\AA\ should be added to the indices of the 
probable mid-K giants 2MASS05284595--6614194, 2MASS05284199--6613457, and 
2MASS05284594--6614138, as well as to the stars in the fainter group.

	The CaT indices of bright giants in NGC 1978 appear to more appropriate 
for solar metallicities, even though NGC 1978 has a sub-solar metallicity. 
Evidence of this comes from the metallicity dependance of K4 and K5 III stars with 
log(g) between 1.0 and 2.0 and [Fe/H] between --0.3 and 0.3 in the Cenarro 
et al. (2001a) database. While there are stars with a later spectral-type 
in NGC 1978, the Cenarro et al. (2001a) database does not contain a large 
number of early M giants that span a range of metallicities -- K4 to K5III are the 
latest spectral-types for which a moderately large number of objects (21) is available. 
Applying a least squares fit to the CaT and [Fe/H] values for these stars yields CaT = 
($3.5 \pm 0.7$) Fe/H $+ 10.2 \pm 0.1$. CaT $\sim 8.8$\AA\ would then be expected for 
K giants with log(g) $\sim 1.5$ in NGC 1978, whereas for solar metallicities 
CaT $\sim 10.2$\AA. The mean CaT found here for stars in NGC 1978 is 
$\sim 10.9\AA$. 

	Random uncertainties in the CaT indices among stars in the bright sample 
were estimated by measuring the pointing-to-pointing scatter in the indices of 
stars that were observed in more than one slit position. The random uncertainty in the 
CaT indices of stars in the bright sample found in this manner
is $\pm 0.7\AA$. For comparison, the $1\sigma$ dispersion in CaT among 
the four M giants is $\pm 0.86$ \AA, with a mean instrumental CaT of 10.65\AA. 
The $1\sigma$ dispersion in the CaT indices among the K giants in the bright group and 
all of the stars in the faint sample is $\pm 1.20$\AA, with 
a mean of 10.26\AA. However, the CaT index for 2MASS05284668--6614262 is markedly lower 
than those of the other stars, and this star also has a smaller $J-K$ color than 
the others. If this star is not considered then the dispersion among suspected K 
giants drops to $\pm 0.67$\AA, with a mean instrumental CaT of 10.61\AA. 

	Ferraro et al. (2006) measure [Fe/H] for individual stars in 
NGC 1978, and find values between --0.25 and --0.44, with uncertainties in individual 
measurements typically between $\pm 0.15 - 0.20$ dex. The dispersion in the [Fe/H] 
measurements is $\pm 0.07$ dex, and this sets an upper limit to 
the metallicity dispersion in the cluster. This dispersion in [Fe/H] 
translates to a dispersion in the CaT values of $\pm 0.2\AA$ using the relation 
between CaT and [Fe/H] discussed in the previous paragraph. Star-to-star metallicity 
differences thus do not contribute significantly to the dispersion in the CaT indices.
These results suggest that random uncertainties in the measurements (1) dominate the 
dispersions in CaT, and (2) can not explain the difference between the expected 
and measured mean CaT indices.

	The mean CaT for the M giants is the same 
within the errors with the mean CaT among the suspected K giants, and this 
is not unexpected. The T$_{eff}$ values compiled in Table 5 of 
Cenarro et al. (2001b) indicate that the difference in T$_{eff}$ 
between mid-K giants and early M dwarfs is only a few hundred K. The relation 
between CaT and T$_{eff}$ is examined in the top panel of Figure 3 of 
Cenarro et al. (2002), and it is evident that CaT does not vary greatly with 
T$_{eff}$ near T$_{eff} \sim 4000$ (i.e. the approximate T$_{eff}$ of 
mid-K giants and M giants). 

	There are strong telluric emission features in the wavelength region 
near the Ca triplet, and uncertainties when subtracting these will 
leave residuals in the spectra. If sky emission 
lines are systematically over-subtracted then the CaT indices will be skewed to 
higher values. However, there is evidence that this is not happening here. If 
sky subtraction errors biased the CaT indices in Table 2 then there should be a trend 
between CaT and magnitude, in the sense of deeper CaT absorption 
towards fainter magnitudes if sky emission features have been over-subtracted.
However, the CaT indices in Table 2 do not show such a 
trend. The mean CaT of suspected K giants in the bright sample ($10.6 \pm 0.2$)
is the same as the mean CaT among the faint sample ($10.7 \pm 0.4$), if the 
star 2MASS 05284668--6614262 -- which has a low $J-K$ color and small CaT 
compared with other stars in the sample -- is not included. Therefore, the 
larger than expected CaT indices are likely not the consequence of a systematic 
offset introduced by uncertainties in sky subtraction.

	The larger-than-expected CaT indices in NGC 1978 are almost certainly 
due in part to surface gravity, as the brightest giants in 
NGC 1978 studied by Ferraro et al. (2006) have log(g) $< 1$, and so have lower 
surface gravities than the giants that were used to find the relation between 
CaT and [Fe/H] derived earlier in this section. However, the relation between 
log(g) and the depth of the Ca triplet in Figure 4 of Cenarro et al. (2002) 
indicates that a change in log(g) of 0.5 dex changes CaT by an 
\AA\ in the low surface gravity regime, and so could explain only part of 
the difference between the observed and expected CaT indices. We note that applying a 
log(g) correction of this size to the expected CaT computed using the relation found 
above yields a CaT index for solar metallicity stars that agrees with that measured 
among giants in NGC 1978. Still, the CaT index from the integrated spectrum 
of NGC 1978 -- discussed in Section 3.2 -- is also stronger (1) than 
predicted by models, and (2) than in the integrated spectrum of NGC 55 
(Section 4.1), which has a similar metallicity and luminosity-weighted age as NGC 
1978. This suggests that the high CaT indices are an intrinsic property of the 
cluster members. Possible causes of the strong Ca lines are discussed in Section 6.

\subsection{Integrated Spectra}

\subsubsection{The influence of bright stars}

	Spectra of NGC 1978 were extracted from each slit pointing by summing light 
over a 40 arcsec wide interval centered on the peak of the cluster light profile. 
The spectrum at each pointing and the mean of these are shown in Figure 4. 
Pointing-to-pointing differences are evident. 
Large differences are associated with the CN band heads, reflecting 
the contribution that 2MASS05284449--6614039 makes to the light from 
P3 and P4, and its negigible contribution to the light from the other pointings. 
2MASS05284449--6614039 accounts for almost half of the 
light from P3 between 0.7 and $0.9\mu$m. The comparisons in Figure 4 
clearly indicate the potential that CN band heads near 7900\AA\ and 9100\AA\ 
have for identifying C stars in integrated light. There are also differences 
between the spectra of P1, P2, and P5. P1 and P5 each contain 
at least one of the bright M giants listed in Table 2, and 
the TiO band head at 7100\AA\ in the integrated spectra of P1 and P5 
is deeper than in the spectrum of P2. 

\begin{figure}
\figurenum{4}
\epsscale{1.0}
\plotone{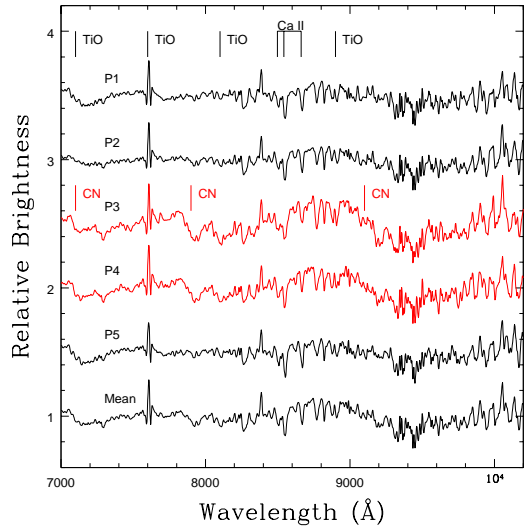}
\caption{Spectra of the five pointings in NGC 1978 and the average of these. 
The spectra of P3 and P4 are shown in red, to signify that the C star 
2MASS05284449--6614039 contributes light to these pointings. Pointing-to-pointing 
differences in spectroscopic properties are seen. 
Differences are clearly evident in the depths of CN bands, which are deepest 
in the spectra of P3 and P4. The spectra of the pointings that do not 
contain signal from the C star show differences in the depths 
of the TiO bands. These comparisons indicate that a spectrum 
of NGC 1978 that is based on a single slit observation would not capture 
a representative SED of the cluster.}
\end{figure}

	Spectra of the five pointings in a 300\AA\ wide interval 
centered on the Ca triplet are compared in the left hand panel of Figure 5. The 
differences between the average spectrum and the individual spectra are shown in the 
right hand panel. Pointing-to-pointing differences on the order of a few percent 
are seen at wavelengths near the Ca triplet. Not surprisingly, 
the largest residuals occur for P3, where the C star makes the largest fractional 
contribution to the total light, and the residuals for this pointing between 8500 and 
8600\AA\ track CN absorption features. The comparisons in Figures 4 and 5
demonstrate that a hypothetical spectrum of NGC 1978 obtained from a single slit 
observation would have difficulty capturing a respresentative SED of the cluster, 
even if the dense central regions of the cluster are sampled.

\begin{figure}
\figurenum{5}
\epsscale{1.0}
\plotone{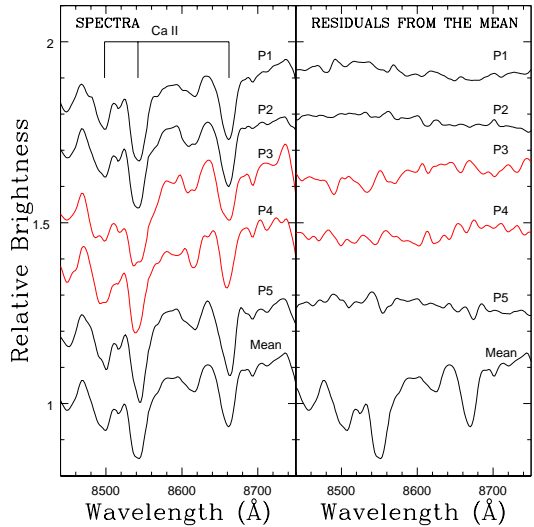}
\caption{Spectra of the five NGC 1978 pointings are shown in the left hand panel, while 
the differences between the average spectrum of all pointings and the 
individual spectra are shown in the right hand panel. The spectra of P3 and P4, where 
the C star 2MASS05284449--6614039 contributes much of the light, are shown in red. 
The structured nature of the P3 residuals between 8500 and 8600\AA\ is due to CN 
absorption features.}
\end{figure}

	The instrumental CaT index of each pointing is 
listed in Table 3. Given the bright nature of NGC 1978, the uncertainties in the 
CaT indices in Table 3 are expected to be smaller than those estimated for individual 
stars in Section 3.1. A surprising result is that -- despite the distinct 
spectroscopic characteristics of C star spectra -- the sizeable contribution 
made by 2MASS05284449--6614039 to the integrated light does not skew the CaT indices 
of P3 and P4 to values that differ from those in the other
pointings. Indeed, the CaT indices of P3 and P4 fall 
in the middle of the range listed in Table 3. The CaT index for P5, 
which contains the brightest M giant, is larger than that of the 
other four pointings, and in Figure 5 it can be seen that the Ca lines 
in the P5 spectrum are visually deeper than in the other pointings. 

\begin{deluxetable}{lc}
\tablecaption{NGC 1978 CaT Indices}
\startdata
\tableline\tableline
Pointing & CaT \\
 & (\AA) \\
\tableline
P1 & 8.74 \\
P2 & 8.23 \\
P3 & 8.31 \\
P4 & 8.32 \\
P5 & 9.78 \\
All P & 8.65 \\
 & \\
P5 (No M star) & 9.16 \\
All P (No M star) & 8.55 \\
 & \\
P3 (No C Star) & 7.96 \\
All P (No C Star) & 8.79 \\
\tableline
\enddata
\end{deluxetable}

	Many highly evolved intermediate age stars are long 
period variables, with light curve amplitudes of a magnitude 
or more. The integrated spectrum of a cluster might then change subtley
as the light output from its brightest members varies with time, and the GMOS 
spectra can be used to explore this matter. The affect that 
variability has on the spectrum of single slit pointings was examined by removing the 
signal of the brightest M giant 2MASS05284364--6613530 \footnote[3]{This star has the 
brightest magnitude between 0.7 and 0.9$\mu$m as 
judged from the GMOS spectra, but does not have the 
brightest $K$ magnitude in Table 2. Highly evolved stars of this nature 
may be photometric variables, and there may also be star-to-star differences 
in $i'$--K color.} from P5, and the signal of the C star 2MASS05284449--6614039 
from P3. The CaT index was measured in the subtracted spectra, and the results 
are listed in Table 3. Removing 2MASS05284364--6613530 reduces the CaT index of P5 by 
0.6\AA, bringing it into better agreement with the CaT values measured in the 
other pointings. Removing 2MASS05284449--6614039 from 
P3 lowers the CaT index by 0.3\AA.

	The removal of light from individual stars will of 
course have a smaller impact on a composite spectrum constructed from many slit 
pointings, as the contribution from individual stars to the total light is diminsihed. 
The C star 2MASS05284449--6614039 accounts for $\sim 17\%$ of the total light 
near 8500\AA\ from all five NGC 1978 pointings, while the M giant 
2MASS05284364--6613530 accounts for $\sim 8\%$ of the total light. 
The results of subtracting 2MASS05284449--6614039 and 2MASS05284364--6613530
from the summed spectrum of all five pointings are examined in Figures 6 
and 7. 

\begin{figure}
\figurenum{6}
\epsscale{1.0}
\plotone{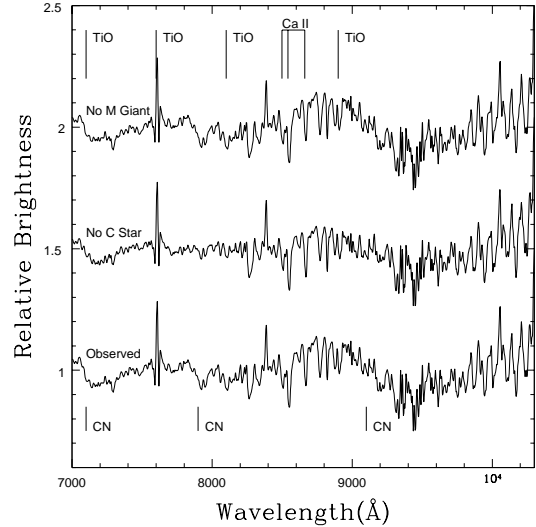}
\caption{Results of subtracting the C star 2MASS05284449--6614039 and the brightest 
M giant 2MASS05284364--6613530 from the mean spectrum of NGC 1978. The mean spectrum 
of all five pointings with no stars removed is also shown for reference. 
Subtracting the brightest M giant has a smaller visual impact on the cluster spectrum 
than removing the C star. This reflects the smaller contribution made by the M giant 
to the total light, coupled with the presence of other bright 
M giants in the area observed with GMOS that have spectral 
signatures similar to those in the 2MASS05284364--6613530 spectrum. Note that 
removing the C star has a noticeable impact on the depth of the CN band heads.}
\end{figure}

\begin{figure}
\figurenum{7}
\epsscale{1.0}
\plotone{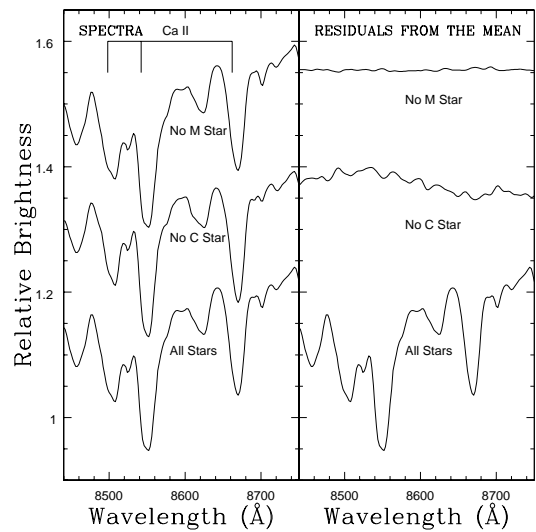}
\caption{Spectra in Figure 6 at wavelengths near the Ca triplet 
are shown in the left hand panel; the differences between the individual 
spectra and the average spectrum with no stars removed are shown in the 
right hand panel. Removing the light from the brightest M giant has only 
a modest impact on the Ca triplet lines, as there are many 
other bright M giants that have similar spectroscopic characteristics 
in the area sampled with GMOS. While removing the signal from the C star changes the 
composite spectrum near the Ca triplet by a few percent, the net impact on 
the CaT index is small.}
\end{figure}

	Given the large fractional contribution that it makes to the NIR light -- 
coupled with the distinct spectroscopic characteristics of C star spectra -- 
it is not surprising that the subtraction of light from 2MASS05284449--6614039 
still has a noticeable impact on the visual appearance of the mean 
spectrum in Figure 6, due largely to the changes wrought on the CN bands. 
These comparisons highlight the potential of using the 7900 and 9100\AA\ CN band heads 
to probe C star content in integrated light. While there is water 
absorption bands between 0.9 and $1.0\mu$m that could complicate CN measurements 
at these wavelengths, the 7900\AA\ band head produces a distinct $\sim 100$\AA\ wide 
feature in the spectrum. 

	In contrast to the CN bands, the depths of the 
Ca triplet, as gauged by the CaT index, does not change by a 
large amount. CaT indices were measured from the mean spectrum with 
2MASS05284449--661403 and 2MASS05284364--6613530 removed, and the results are 
listed in Table 3. Removing light from 2MASS05284449--661403 changes the CaT index 
of the summed spectrum by $\sim 0.1\AA$, and a similar difference occurs 
when 2MASS05284364--6613530 is removed. These modest changes suggest that 
-- when a large fraction of the light from a cluster is considered -- the CaT index 
is a robust probe of the stellar content of intermediate age clusters like 
NGC 1978, even if C stars contribute a large fraction of the light. 

\subsubsection{Comparisons with models}

	The summed NGC 1978 spectrum is compared with model spectra of 
simple stellar populations (SSPs) in Figures 8 and 9. 
The models are from the EMILES (Rock et al. 
2016) compilation, and use the evolutionary tracks of Pietrinferni et al. 
(2004). The models assume [M/H] $\sim -0.35$ with a scaled-solar 
chemical mix and a Chabrier (2003) mass function (MF). 
The model spectra were re-sampled and smoothed to match the 
pixel sampling and wavelength resolution of the GMOS spectra, and 
have had the continuum removed.

\begin{figure}
\figurenum{8}
\epsscale{1.0}
\plotone{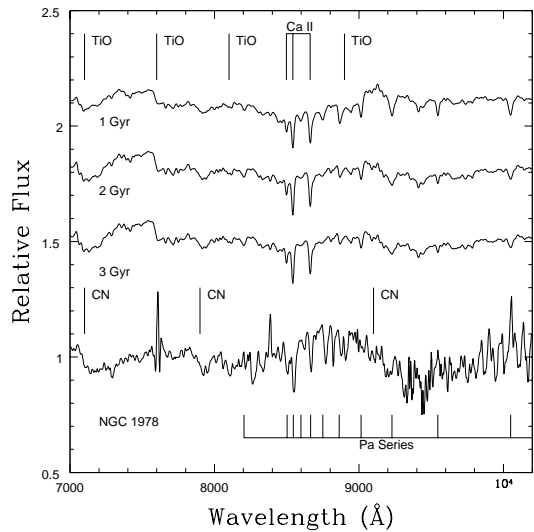}
\caption{Comparisons with [Fe/H] = --0.35 EMILES model spectra. The models 
use the scaled-solar Pietrinferni et al. (2004) isochrones with a 
Chabrier (2003) MF. The models have been processed 
to match the spectral resolution and sampling of the GMOS observations, 
including continuum removal. Based on overall appearance, 
the NGC 1978 spectrum agrees best with the 2 
and 3 Gyr models, although there are difficulties matching the depths of the Ca II 
lines (see text). The 7900\AA\ band head is deeper in the NGC 1978 spectrum 
than in the models, although the significance of this is low given the 
large contribution made by only one C star to the NGC 1978 spectrum.}
\end{figure}

\begin{figure}
\figurenum{9}
\epsscale{1.0}
\plotone{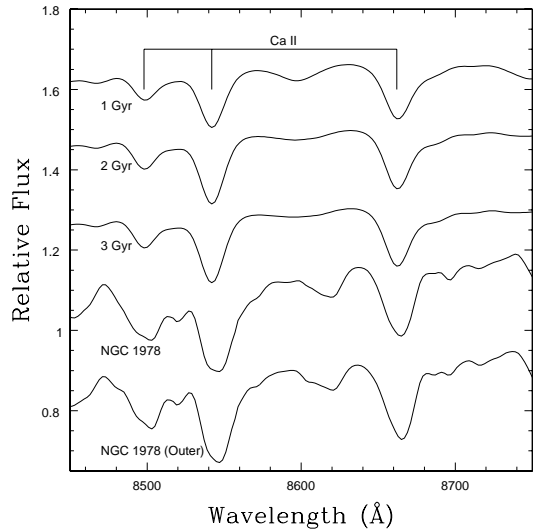}
\caption{Same as Figure 8, but showing wavelengths near the Ca triplet. The spectrum 
that results if light from $\pm 10$ arcsec of the cluster center -- where there 
is a red cusp in the NIR light profile (see text) -- is excluded 
is shown at the bottom. The depths of the Ca lines do not 
change significantly if light from the cluster center 
is excluded. The depths of individual Ca lines in the models are shallower than 
in the NGC 1978 spectra.}
\end{figure}

	The overall appearance of the NGC 1978 
spectrum is matched best by the 2 and 3 Gyr models, in agreement with the 
age estimated by Mucciarelli et al. (2007) from the CMD. Still, there 
is mixed agreement matching the depths of individual features. 
Absorption lines from the Paschen series of hydrogen are 
prominent in the 1 Gyr model, and are much weaker in the 2 and 3 Gyr models. 
There is no evidence of Pa absorption in the NGC 1978 spectrum, although 
the S/N ratio of the NGC 1978 spectrum is poor at wavelengths where the 
strongest Pa lines are expected. The models in Figure 
8 have CN absorption at 7900\AA\ , and the depth of this feature is 
weaker than observed in NGC 1978. However, the significance of this difference 
is low, given that only one C star is sampled in the NGC 1978 spectrum. 

	As for the Ca triplet lines, the agreement 
between the models and the NGC 1978 spectrum 
in Figure 9 is far from perfect, as the Ca lines in the models are shallower 
than in the cluster spectra. The CaT indices measured from the models 
are 6.4 (1 Gyr), 6.8 (2 Gyr), and 6.7\AA\ (3 Gyr), whereas the CaT index 
for the composite NGC 1978 spectrum is 8.65\AA. 
The experiments discussed in the previous section -- in 
which the brightest M giant or C star are subtracted from the spectrum -- indicate 
that the larger than expected CaT index in NGC 1978 is not a 
stochastic effect due to a single luminous star, as the 
removal of the brightest stars changes CaT by only 0.1\AA.

	One possible explanation for the difference in 
Ca line strengths is that the models might not reproduce the number and/or types
of bright giants in NGC 1978. However, we are loath to attribute the failure to 
match the Ca II lines to deficiencies in the model physics. 
EMILES model spectra are available for the Padova00 isochrones (Girardi et al. 
2000), and these differ from those constructed from the Pietrinferni et al. (2004) 
isochrones by only a few percent near the Ca triplet. 
Perhaps more importantly, in Section 4 it is shown that the model spectra 
reproduce the Ca lines in the spectrum of the dwarf irregular galaxy NGC 55 which -- 
while a composite stellar system -- has a characteristic age and 
metallicity that is similar to that of NGC 1978. This agreement also rules out 
deficiencies in the stellar library used to generate the model spectra. 
Finally, in the previous section it was shown that the Ca indices in 
individual NGC 1978 giants appear to be larger than expected.

	An interesting test would be to compare the luminosity function of 
NGC 1978 with that predicted by the models. Unfortunately, we are not aware of a 
published catalogue of complete photometric measurements in NGC 1978 that 
would allow such a comparison to be made. Still, there are mechanisms that might 
produce an excess fraction of low surface gravity stars in the region sampled by 
GMOS. One such mechanism is mass segregation, in the sense that the most massive 
stars occur preferentially near the cluster center, which contributes much 
of the light to the GMOS spectrum. Mass segregation of this nature can set in 
naturally during the early dynamical evolution of star clusters (e.g. Spera et al. 
2016).

	As it turns out, the stellar content of NGC 1978 is not 
uniformly mixed. Evidence for this comes from $J$ and $K$ surface 
brightness measurements obtained from 2MASS All Sky Survey images, which were 
downloaded from the Interactive 2MASS Image Service 
\footnote[4]{http://irsa.ipac.caltech.edu/applications/2MASS/IM/interactive.html}. 
Isophotal photometry was performed with the STSDAS routine 
$ellipse$, which uses the procedures described by Jedrzejewski (1987). 
Background sky levels were measured at large offsets from the 
cluster. Bright resolved stars were suppressed by smoothing the images with an $11 
\times 11$ arcsec median filter, although residual light from the brightest 
stars persists at a low level.

	The $J$ and $K$ surface brightness profiles of NGC 1978 are shown 
in the top panel of Figure 10, while the $J-K$ color profile is 
shown in the lower panel. The radial measurements are 
along the major axis. A red cusp dominates the color profile 
within $< 10$ arcsec of the cluster center. The flat appearance of the cusp at 
small radii is an artifact of the median filter that was applied to the images. 
At larger radii -- and ignoring the two points between 
50 and 60 arcsec -- there is a tendency for $J-K$ to drop gradually with increasing 
radius. 

\begin{figure}
\figurenum{10}
\epsscale{1.0}
\plotone{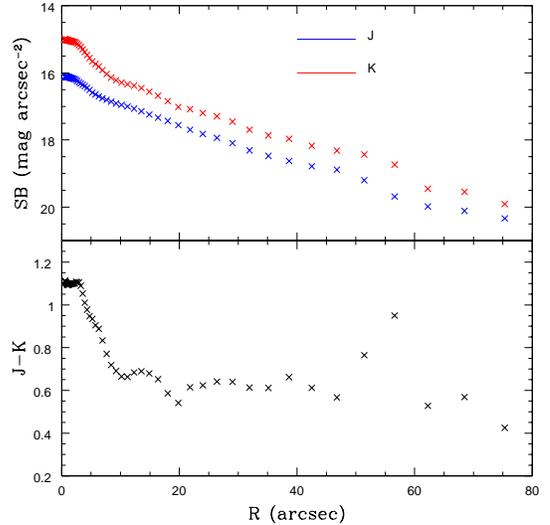}
\caption{Isophotal photometry of NGC 1978 obtained from 2MASS All Sky Survey images. 
Radii are measured along the major axis, which parallels 
the orientation of the GMOS slit. The red cusp within 10 arcsec 
of the cluster center is due to a central concentration of 
luminous giants. Radial variations in stellar content are not restricted 
to the central regions of NGC 1978; with the exception of the two measurements 
between 50 and 60 arcsec, there is a gradual -- but steady -- decline in $J-K$ from 
10 arcsec to 80 arcsec.}
\end{figure}

	The red cusp is not due to the C star 
2MASS05284449--661403, which is the dominant cluster member 
in the NIR. To demonstrate this, light from that star was 
subtracted from the 2MASS images, and $ellipse$ was run again. The cusp remains 
after the C star is removed, although the central $J-K$ color is lowered 
by $\sim 0.3$ magnitudes. 

	The spectrum of NGC 1978 does not change significantly if light from the red 
cusp is omitted. To demonstrate this, a spectrum of NGC 1978 was constructed 
by excluding light within $\pm 10$ arcsec of the cluster center, and 
the result is shown at the bottom of Figure 9. 
The Ca lines in the resulting spectrum have depths that are similar to those 
in the spectrum based on the entire area sampled with GMOS. These results 
suggest that the deeper than expected Ca triplet lines 
in the integrated NGC 1978 spectrum are intrinsic to the 
cluster members, and are not due to the concentration of the brightest cluster members 
near the cluster center. 

\section{RESULTS: NGC 55}

\subsection{NGC 55: Integrated Spectra}

\subsubsection{Properties of the NGC 55 spectrum}

	Davidge (2005) found that the peak $i'$ brightnesses of the most 
luminous AGB stars in NGC 55 changes with distance from the disk plane, 
in the sense that the peak brightness decreases with increasing 
extraplanar distance. The GMOS slit was oriented to be perpendicular to the 
major axis of NGC 55 for these observations, and so gradients in the strengths of 
spectral features might be expected in the GMOS spectra. 
Spectra that sample three angular intervals were thus extracted. 
One interval samples the light within $\pm 14$ arcsec of the disk plane, where the 
disk plane is defined as the midpoint of the light profile through the disk after 
collapsing the GMOS spectra along the dispersion direction between $0.7\mu$m and 
$0.9\mu$m. The other two intervals cover angular offsets 14--28 arcsec and 28--44 
arcsec from the disk plane. Spectra extracted on each side of the major axis were 
averaged together. The three angular intervals sample comparable 
integrated magnitudes, and so the extracted spectra have similar S/N ratios.
Each angular interval samples a region with M$_K \sim -13.5$, which is 
comparable to the light from an intermediate age LMC cluster. 

	The extracted spectra are compared in Figures 11 and 12. 
Subtle changes with distance from the disk plane are seen 
in the strengths of many absorption and emission features. The Ca triplet and 
the TiO band head near $0.71\mu$m tend to weaken with distance from the disk 
plane. The CN band head at 7900\AA , which is not strong but is still present 
in the NGC 55 spectra, may weaken with distance from the disk plane. 
The Paschen discontinuity near $0.82\mu$m is evident in all three spectra. 

\begin{figure}
\figurenum{11}
\epsscale{1.0}
\plotone{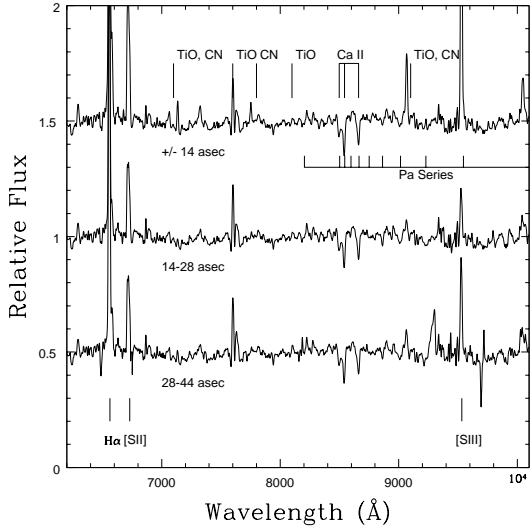}
\caption{Spectra of NGC 55 in three angular intervals, with offsets measured 
along the minor axis from the midpoint in the light profile constructed from the 
GMOS spectra. The Ca triplet and TiO band heads weaken with 
increasing distance from the disk plane, indicating 
that the mix of stars that dominate the light at these wavelengths also changes. 
The CN band head near 7900\AA\ produces a weak break in the spectrum. 
The strengths of emission lines differ from spectrum-to-spectrum, indicating that the 
sources of ionzing radiation are not well mixed throughout the area sampled by GMOS.
The strongest emission lines are due to the H$\alpha +$ S[II] blend, as well 
as [SII] and [SIII].}
\end{figure}

\begin{figure}
\figurenum{12}
\epsscale{1.0}
\plotone{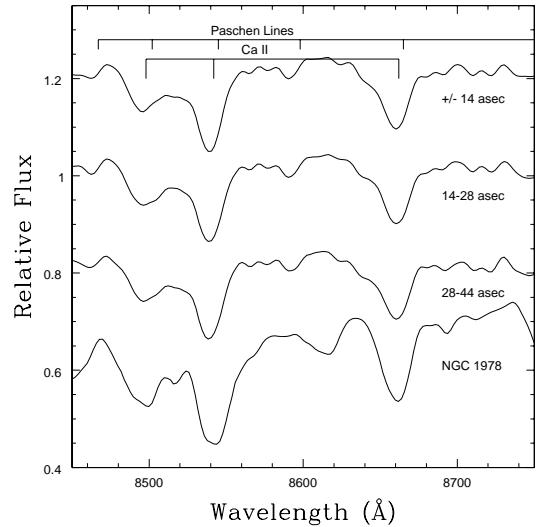}
\caption{Same as Figure 11, but at wavelengths near the Ca triplet. 
The summed spectrum of NGC 1978 is also shown. The depths 
of the Ca triplet and high order Pa lines, as gauged with the CaT and PaT 
indices, vary with distance from the disk plane. The CaT and PaT indices are 
consistent with a population having a light-averaged 
age $1-2$ Gyr and [Fe/H] $= -0.35$ (see text). The Ca lines are deeper in 
NGC 1978 than in NGC 55, even though bright stars in 
NGC 55 and NGC 1978 have similar [Fe/H], and the effective age of NGC 55 
is comparable to that of NGC 1978.} 
\end{figure}

	Prominent emission lines are seen in all three spectra. 
Large spectrum-to-spectrum variations in H$\alpha$ and 
[SIII] are evident, as well as in the strengths of lower order Pa lines. 
The variations in emission line strengths 
indicate that the hot stars that are the likely source of ionizating radiation 
are not uniformly mixed throughout the area sampled by GMOS. 
This is not unexpected, as line emission in NGC 55 has a 
complex morphology (Ferguson et al. 1996), with bubbles and jets 
protruding from the disk plane. Ferguson et al. (1996) conclude 
that much of the H$\alpha$ emission is associated with two 
kpc-scale star-forming complexes located near the center of the galaxy. 

	Veiling of absorption lines by continuum emission is not expected to be 
significant in these data. Reines et al. (2010) use Starburst99 
(Leitherer et al. 1999) models to examine the contribution made by 
continuum emission to the total signal from SSPs. The equivalent width of 
H$\alpha$ emission in NGC 55 is consistent with an age of $\sim 10$ Myr, 
and the nebular continuum contributes only a small fraction of the total light 
for such ages (Figure 8 of Reines et al. 2010). 
That the Ca lines show only modest changes in strength in the three intervals 
examined in NGC 55, despite large spectrum-to-spectrum variations in 
the strength of H$\alpha$, further suggests that veiling by continuum emission 
is not significant. 

	Cenarro et al. (2001a) define an index -- PaT -- that 
measures the depths of Paschen lines at wavelength near the Ca triplet, 
and this index has been measured in the NGC 55 spectra. 
Instrumental CaT and PaT indices for the NGC 55 spectra are listed 
in Table 4. The CaT index becomes smaller (i.e. decreasing equivalent 
widths of Ca lines) with increasing distance from 
the disk plane. The CaT indices in NGC 55 are more-or-less consistent 
with those of the [Fe/H] $= -0.35$ models discussed in Section 3.2.2.
The PaT indices are modest in size, and decrease with increasing 
distance from the disk plane, transitioning from weak absorption to weak emission.

\begin{deluxetable}{lcc}
\tablecaption{Instrumental CaT Indices for NGC 55}
\startdata
\tableline\tableline
Point & CaT & PaT \\
 & (\AA) & (\AA) \\
\tableline
0 -- 14 asec & 6.76 & 0.3 \\
14 -- 28 asec & 6.41 & --0.1 \\
28 -- 44 asec & 5.97 & --0.4 \\\\
cluster 1 & 12.21 & 2.7 \\
\tableline
\enddata
\end{deluxetable}

	Line emission is expected to be modest for the Paschen series transitions 
that are targeted by the PaT index, although the PaT indices in the outer two 
intervals suggest that line emission is present.
Assuming Case B recombination and an electron temperature 
of 10$^4$ K, then emission from the n=12 and n=14 transitions that are examined by 
the PaT index will have intensities that are only $0.2 - 0.3\%$ those of H$\alpha$ 
(Brocklehurst 1971). The equivalent width of H$\alpha$ in the center spectrum is 
$\sim 50$\AA , and $\sim 20$\AA\ in the 28 -- 44 arcsec spectrum. Therefore, line 
emission in the n=12 and n=14 lines is expected to have an equivalent width $\leq 
0.1$\AA\ assuming Case B recombination. This is roughly consistent with the PaT 
indices measured in the 14--28 and 28--44 arcsec spectra, thereby suggesting that Pa 
absorption due to the n=12 and n=14 transitions is modest in these angular intervals.

\subsubsection{Comparisons with NGC 1978 and model spectra}

	The metallicity of young stars in NGC 55 is similar to that in NGC 1978 (e.g. 
Ferraro et al. 2006; Patrick et al. 2017 and references therein). The integrated 
spectrum of NGC 1978 is compared with the NGC 55 spectra in Figure 12, and 
the Ca absorption lines in NGC 55 are weaker than in NGC 1978. This difference 
occurs even though much of the light from NGC 55 at red and NIR wavelengths 
originates from intermediate age stars that have an age that is 
not greatly different from that of NGC 1978 (see below).

	The characteristic age of the NGC 55 spectrum is explored 
in Figures 13 and 14, where the mean NGC 55 spectrum -- constructed by combining 
the three spectra in Figure 11 -- is compared with SSP model spectra. 
The models use the scaled-solar abundance Pietrinferni et al. et al. (2004) 
isochrones. [Fe/H] $=-0.35$ has been adopted with a Chabrier MF.
As in Section 3, the model spectra were processed 
to match the spectral resolution and sampling of the GMOS data and have had 
the continuum removed. The metallicity of the 10 Gyr 
population in Figures 13 and 14 is the same as that of the younger 
populations, and so a flat age-metallicity relation has been assumed. 

\begin{figure}
\figurenum{13}
\epsscale{1.0}
\plotone{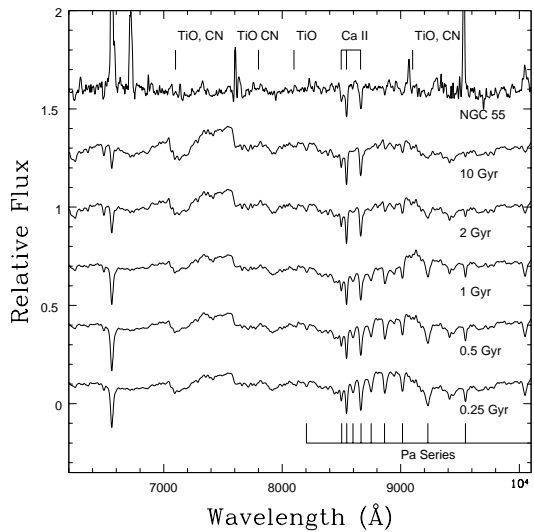}
\caption{Mean spectrum of NGC 55, constructed by combining the three spectra in 
Figure 11. Model spectra of SSPs with a metallicity --0.35 and a Chabrier MF 
constructed with the scaled-solar Pietrinferni et al. (2004) isochrones are also 
shown. While the emission lines indicate that hot young 
main sequence stars are present, the absence of deep Pa absorption lines 
indicates that the bulk of the light in the absorption spectrum at these wavelengths 
originates from a population that has an age $\geq 1$ Gyr. 
The weak TiO band heads at 7100 and 7600\AA\ indicate that much 
of the light originates from stars with an age $\leq 2$ Gyr.}
\end{figure}

\begin{figure}
\figurenum{14}
\epsscale{1.0}
\plotone{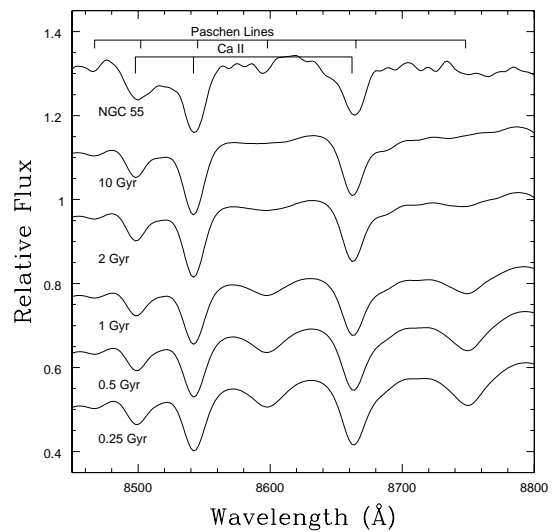}
\caption{Same as Figure 13, but showing wavelengths near the Ca 
triplet. The depth of the Ca triplet in the NGC 55 spectrum comes closest to 
matching that in the 2 and 3 Gyr models. The n=12 and n=14 Paschen lines near 8750 
and 8650\AA\ are deepest in models with ages $\leq 1$ Gyr. That these lines are 
not seen in the NGC 55 spectrum argues for a characteristic age $\geq 1$ Gyr.} 
\end{figure}

	The best overall match between the models and the observed 
spectra in Figures 13 and 14 is found with the 1 or 2 Gyr models. The depths 
of the Ca absorption lines in the NGC 55 spectrum in Figure 14 match 
those in the 2 and 10 Gyr models, and are deeper than in the younger models. 
The CaT index in the $\pm 14$ arcsec NGC 55 spectrum matches that in the 2 Gyr model. 
Still, the 2 and 10 Gyr models have much stronger 
TiO band heads at 7100 and 7600\AA\ than are observed, suggesting that the 
characteristic age of NGC 55 at red wavelengths is not `old'. An upper limit of 2 
Gyr is thus set for the characteristic age of the red spectrum of NGC 55.

	A lower limit to the characteristic age of NGC 55 is set using 
the depths of the Paschen lines. The models with ages $\leq 1$ Gyr in Figure 14 
have moderately strong Pa absorption features near 8650\AA\ (n=14 transition) and 
8750\AA\ (n=12 transition), and these are abscent in the NGC 55 spectrum. 
The absence of Pa lines in Figure 14 then argues for an age $\geq 1$ Gyr. With 
an effective age 1 -- 2 Gyr for the red/NIR light from NGC 55 
then the CN band head near 7900\AA\ may have a contribution from C stars. However, 
the shallow nature of this feature indicates that C stars are not present 
in the numbers predicted by the models.

\subsubsection{Low mass main sequence stars in NGC 55}

	The main focus of this paper is the influence 
that the brightest stars in a system have on the red/NIR spectrum. However, 
absorption from Na I near 8190\AA\ and FeH near 9914\AA\ 
provide insights into the low mass end of the MF. The good quantum 
efficiency of the Hamamatsu CCDs near $1\mu$m makes these data 
particularly well-suited to search for FeH absorption. The portions of the 
mean NGC 55 spectrum that sample these features are shown in Figure 15. The spectrum of 
the M5V star Gliese 51 from Rayner et al. (2009)  -- smoothed to match the 
spectral resolution of the NGC 55 observations -- is also shown, as are model spectra 
of 2 and 10 Gyr populations that were constructed with a Chabrier MF. 

\begin{figure}
\figurenum{15}
\epsscale{1.0}
\plotone{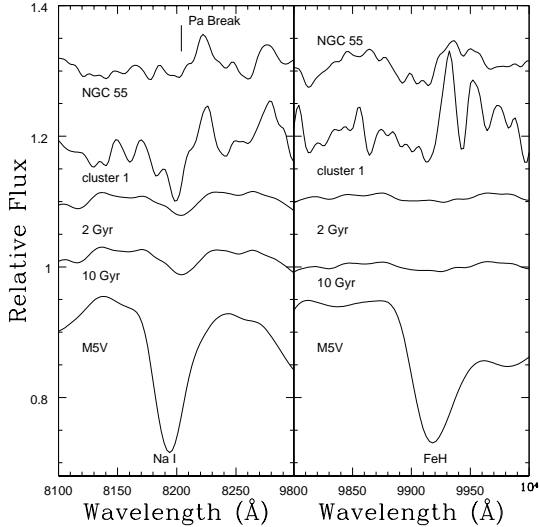}
\caption{Spectra of NGC 55 and cluster 1 near Na I 8190 and FeH 9914. 
The spectrum of the M5V star Gliese 51 from Rayner et al. (2009), smoothed 
to match the resolution of the GMOS data, is also shown, along with SSP models 
constructed with a Chabrier MF. Na I 8190 is not 
detected in NGC 55, although the Paschen break at 8204\AA\ complicates 
efforts to detect faint absorption features near this wavelength. 
In contrast, absorption with the same wavelength as Na I is present in the 
cluster 1 spectrum, although this may be interstellar in origin 
(see text). Absorption near 9900\AA\ in the NGC 55 and cluster 1 spectra 
coincides with part of the FeH band, although the detection of the red edge 
of this feature is complicated by possible line emission.}
\end{figure}

	The depths of the Na I and FeH features in the models have a mild 
sensitivity to age among old and intermediate age systems. 
Na I is deeper in the 10 Gyr model, and this 
reflects the diminishing (but still substantial) contribution that red 
giants make to the red light as one moves to ages older than 2 Gyr. Na I absorption 
has a depth of a few percent in the models. There is no evidence 
for Na I absorption in the NGC 55 spectrum, although 
the Paschen break at 8204\AA\ warps the local continuum, 
complicating the detection of faint absorption features at nearby wavelengths.

	The situation may be different for FeH. The models show that FeH produces 
a broad absorption feature with a depth of $\sim 1\%$. The 
shallow nature of the FeH feature in the models highlights the challenges 
of detecting FeH in systems that do not have a 
bottom-heavy MF. Still, a wide absorption feature near 9900\AA\ is seen in the NGC 55 
spectrum, the short wavelength onset of which matches 
that of FeH in the M5V spectrum. However, this feature is narrower in NGC 55 
than in the M5V spectrum. The red part of FeH absorption in NGC 55 may be affected by 
emission near 9930\AA , which could prevent the detection of the red edge of the FeH 
feature. The source of this emission is not known. Given the inability to match the 
full wavelength range of the absorption feature, coupled with the 
possible presence of emission lines that could alter the local continuum 
at these wavelengths, then it is premature to identify the absorption 
near 9900\AA\ in the NGC 55 spectrum with FeH. 

\subsection{NGC 55: The Spectrum of Cluster 1}

	The GMOS slit passes through the star-forming complex that Davidge (2005) 
named cluster 1. The light profile of the part of cluster 1 that is sampled with GMOS 
has a FWHM  $\sim 5.4$ arcsec at wavelengths between 0.7 and $0.9\mu$m, and a spectrum 
was constructed by summing the light within this angular interval. 
Light from the main body of NGC 55 contributes significantly to cluster 1, 
and so a local sky spectrum was constructed by combining signal at 
the points where the cluster light profile blends into the main body 
of the galaxy. The local sky spectrum was then subtracted 
from the extracted spectrum to obtain the final cluster 1 
spectrum that is discussed here.

	The spectrum of cluster 1 is shown in Figures 16 and 17. Many of the 
strong emission lines in the NGC 55 spectrum are either removed or 
over-corrected by the subtraction of the local sky. The lines of H$\alpha$, 
[SII] and [SIII] are over-corrected, and these features have been 
clipped at a level that is $20\%$ lower than the continuum in Figure 16 
to prevent interfering with the other spectra shown in the figure. That emission lines 
are stronger outside of cluster 1 suggests that 
the objects that are the dominant drivers of line emission 
are either (1) not concentrated in this part of cluster 1, but 
instead are located outside of the region sampled by GMOS, or (2) are present in 
cluster 1, but have created a bubble in the local ISM that girds the part of 
cluster 1 that is sampled by GMOS.

\begin{figure}
\figurenum{16}
\epsscale{1.0}
\plotone{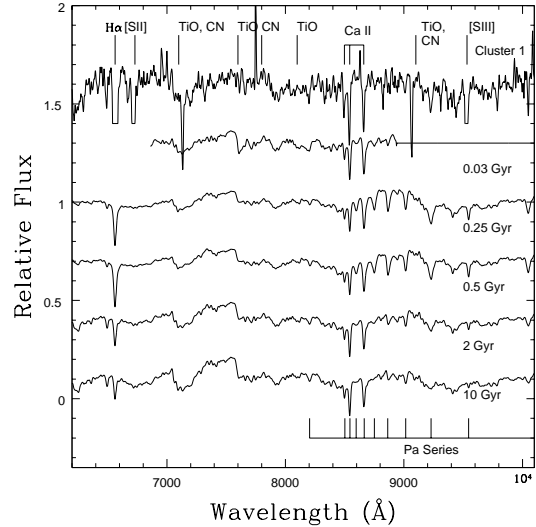}
\caption{Spectrum of NGC 55 cluster 1 compared with 
[Fe/H] $=-0.35$ models from the EMILES compilation that assume a 
Chabrier MF. The 0.03 Gyr model does not extend past $0.9\mu$m. Many of 
the emission features seen in the NGC 55 spectra in Figure 13 
are suppressed or over-corrected in the cluster 1 spectrum. The over-subtracted 
signatures of H$\alpha$, [SII], and [SIII] have been clipped at 20\% below the 
continuum level to prevent interfering with other spectra. The absorption 
spectrum of cluster 1 is dominated by features that originate in the photospheres of 
cool stars. The deep, sharp Ca lines suggest either a metallicity that is higher 
than the value adopted for NGC 55, or that there is a large contribution from 
bright RSGs. The CN band head near 7900\AA\ is also seen. That this 
feature is deeper than in any of the models suggests that some of the light 
from cluster 1 at these wavelengths comes from C stars.}
\end{figure}

\begin{figure}
\figurenum{17}
\epsscale{1.0}
\plotone{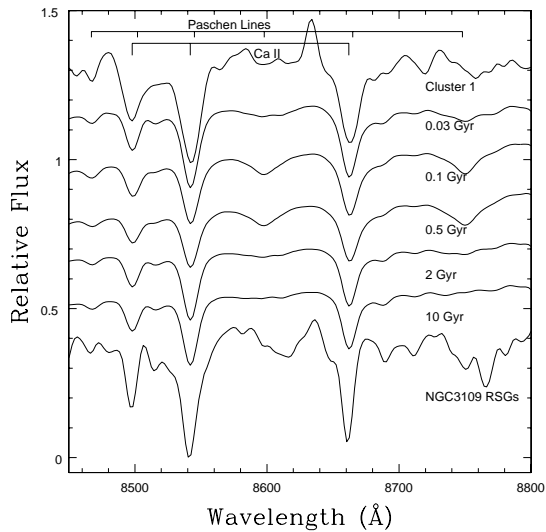}
\caption{Same as Figure 16, but showing wavelengths near the Ca 
triplet. The mean spectrum of the three RSGs detected in NGC 3109 is also 
shown. The cluster 1 spectrum is matched best by 
the 30 Myr model, although the Ca triplet lines 
in the cluster 1 spectrum are deeper and sharper than in the models, with 
strengths that approach those in the composite NGC 3109 RSG spectrum. 
The Pa absorption lines in this part of the spectrum are prominent in the 0.1 
and 0.5 Gyr models -- these lines are weak or not seen in the cluster 1 spectrum. 
These differences between the cluster 1 and model spectra suggest that 
the stars that dominate the Ca triplet in cluster 1 have an age $< 30$ Myr. 
An older component with an age $> 1$ Gyr may also be present 
based on the CN band head near 7900\AA , although this feature also appears in 
the 30 Myr model.}
\end{figure}

	The GMOS spectra support the first possibility. The deep 
Ca triplet lines in the cluster 1 spectrum indicate that it 
is old enough for luminous red stars to have developed, pointing to a 
significant contribution from stars with an age $> 8$ Myr. This is an age 
where the luminous blue stars that power line emission disappear. 

	The dominant light contribution from cluster 1 appears to come 
from a population with an age of no more than a few tens of Myr. This is 
because the Ca lines in the cluster 1 spectrum in Figure 17 are sharp and deep, 
as expected if a large fraction of the light comes from RSGs. The mean spectrum 
of RSGs extracted from the NGC 3109 observations -- discussed in the 
next section -- is also shown in Figure 17. The NGC 3109 stars are probably 
$\sim 0.3$ dex more metal-poor than their NGC 55 counterparts.
Still, the Ca lines in the spectrum of these RSGs are only slightly deeper than 
those in the cluster 1 spectrum

	EMILES SSP model spectra are compared with the 
cluster 1 spectrum in Figures 16 and 17. The models 
assume [M/H] $\sim -0.35$ with scaled-solar abundances and a Chabrier 
MF. The TiO bands near 7100\AA\ and 7600\AA\ both match those 
in the 30 Myr model, and the shallow depth of the TiO band head near 
7600\AA\ is consistent with an age $\leq 0.5$ Gyr. 
While the 30 Myr model comes closest to matching the 
cluster 1 observations, the Ca lines are deeper in the 
cluster 1 spectrum than in that model, suggesting that cluster 1 has 
younger, brighter, and cooler RSGs than in the model. While a metallicity in 
excess of that used for the models could explain stronger 
Ca triplet lines, such an interpretation is unlikely for cluster 1 given 
the metallicities measured for young stars in NGC 55 (Patrick et al. 2017 and 
references therein). We thus suspect that cluster 1 contains a large 
contribution from stars with ages between 10 and 30 Myr. The EMILES 
compilation does not contain models with such ages.

	There is a break in the cluster 1 spectrum that coincides with the CN band head 
near 7900\AA . This feature is seen in all of the SSP models in Figure 16, it 
is deeper in cluster 1 than in any of the models. 
Other CN band heads occur near 7100\AA\ and 9100\AA. 
However, these coincide with TiO band heads, and so features at these 
wavelengths in the cluster 1 spectrum can not be linked unambiguously to CN. 
While the depth of the 7900\AA\ band head suggests 
that cluster 1 may contain stars with ages older than a few hundred Myr, 
the CN band head is deepest in the 30 Myr model, suggesting that the CN feature may 
originate in RSGs. If cluster 1 had an age of a few hundred Myr then the models 
indicate that there would be deep Pa absorption lines that originate from bright stars 
near the main sequence turn-off. Pa absorption lines are not 
obvious in the cluster 1 spectrum. While there is a moderately 
deep line near 9200\AA\ that coincides with the expected location of a Pa line, 
it is the only feature that coincides with Pa absorption.

	The cluster 1 spectrum near Na I 8190 and FeH 9914 is examined 
in Figure 15. An absorption feature is seen near 8190\AA\ 
that coincides with Na I, and it should be recalled that a similar feature is not 
present in the NGC 55 spectrum (Section 4.1). If due to photospheric Na I 
then this feature indicates that cluster 1 contains at least as many 
low mass stars as predicted by the Chabrier (2003) MF. 
However, if much of the light in cluster 1 comes from stars with an age $\leq 30$ Myr 
then stars with masses of a few tenths solar or less would not have had time to 
relax onto the main sequence (e.g. Siess et al. 2000), and so would not have the high 
surface gravities that are characteristic of deep Na I absorption. Given the age 
of cluster 1 then the Na absorption might be interstellar in origin, originating in 
gas along the cluster 1 line of sight.

	As for the wavelength region near FeH, a feature that is similar 
to that found in the NGC 55 spectrum is also seen in the cluster 1 spectrum. 
However, the same caveats that were stated for the NGC 55 spectrum apply here, 
in that the match in wavelength coverage with FeH is not good. Also, if stars with 
masses near $0.1$ M$_{\odot}$ have not relaxed on to the main sequence then the 
surface gravities may not be conducive to the formation of hydrides.

\section{RESULTS: NGC 3109}

\subsection{NGC 3109: Stellar Spectra}

	Four distinct bright stars are located in the areas in NGC 3109 that were 
sampled by GMOS. None are obvious blends, although contamination from 
faint stars might go undetected. The spectra of these stars are shown in Figure 
18. Two are in the 2MASS point source catalog, and their NIR 
brightnesses and colors are listed in Table 5. These stars are near the faint limit 
of the 2MASS survey, and are in moderately crowded fields. Therefore, 
the 2MASS photometry is subject to sizeable uncertainties. The 
2MASS naming convention has been adopted to show the approximate co-ordinates 
of the other two stars.

\begin{figure}
\figurenum{18}
\epsscale{1.0}
\plotone{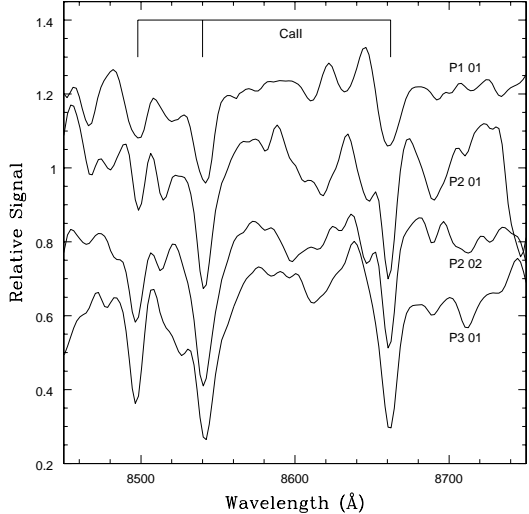}
\caption{Spectra of bright stars in NGC 3109. 
The Ca lines in three of the spectra are deep and 
sharp, as expected for RSGs. These stars also have 
radial velocities that are consistent with membership in NGC 3109. 
In contrast, the Ca lines in the P1-01 spectrum are weaker and broader than in 
the other spectra. P1-01 has a radial velocity v$_r \sim 0$ km sec$^{-1}$, and so 
is a Galactic foreground object.}
\end{figure}

\begin{deluxetable}{clccc}
\tablecaption{Stars in NGC 3109}
\startdata
\tableline\tableline
GMOS ID & 2MASS ID \tablenotemark{a} & $K$ & $H-K$ & $J-K$ \\
 & (RA$+$Dec) & & & \\
\tableline
P1-01 & (10031250--2608558) & -- & -- & -- \\
P2-01 & 10030403--2609027 & 15.630 & 0.121 & 0.892 \\
P2-02 & (10030430--2608403) & -- & -- & -- \\
P3-01 & 10030205--2608575 & 15.782 & 0.179 & 0.781 \\
\tableline
\enddata
\tablenotetext{a}{Entries in brackets indicate stars that are not in the 2MASS PSC. 
The numbers in brackets show RA and Declination, using the 2MASS naming scheme.}
\end{deluxetable}

	Three of the stars are members of NGC 3109, while another is a 
foreground Galactic star. The depths and shapes of the Ca triplet lines in the spectra 
of P2-01, P2-02, and P3-01 are similar, indicating comparable spectral-types. 
These stars have radial velocities $\sim 400$ km sec$^{-1}$, which is 
consistent with membership in NGC 3109. Hence, these stars are RSGs in NGC 3109. The 
near-infrared colors of P2-01 and P3-01 are consistent with this conclusion. 
The mean spectrum of these three stars is shown in Figure 17.
P1-01 has spectroscopic properties that differ from the others, in the sense 
that the Ca absorption lines in the P1-01 spectrum are shallower and broader. 
The radial velocity of P1-01 is $\sim 0$ km sec$^{-1}$, indicating that it is not 
a member of NGC 3109, but instead is in the Galactic foreground.

\subsection{NGC 3109: Integrated Spectra}

	The spectrum of NGC 3109 was constructed by averaging 
signal at each pointing in a $\pm 45$ arcsec wide interval along the slit. 
The center of the extraction window was the 
midpoint of the light profile that was constructed by collapsing the spectra between 
0.7 and $0.9\mu$m. The NGC 3109 spectra have a much lower S/N ratio than 
those of NGC 55 due to the low surface brightness of the galaxy. 
Hence, the spectra of the three pointings were combined to boost the S/N ratio. 
Even then, the S/N ratio of the combined spectrum is poor at most 
wavelengths, and there is substantial noise 
at wavelengths dominated by sky emission lines. In recognition of this, the current 
investigation is restricted to the two wavelength regions 
shown in Figure 19. The top panel covers wavelengths near the band head of 
TiO at 7100\AA. Wavelengths near this feature are free of 
telluric emission lines, while the TiO band head itself is 
a gauge of M giant content and is an age diagnostic 
(see below). The lower panel shows the region near the Ca 
triplet. This wavelength interval was selected for its astrophysical importance, 
although residuals from sky emission lines chop up much of the spectrum 
near these features.

\begin{figure}
\figurenum{19}
\epsscale{1.0}
\plotone{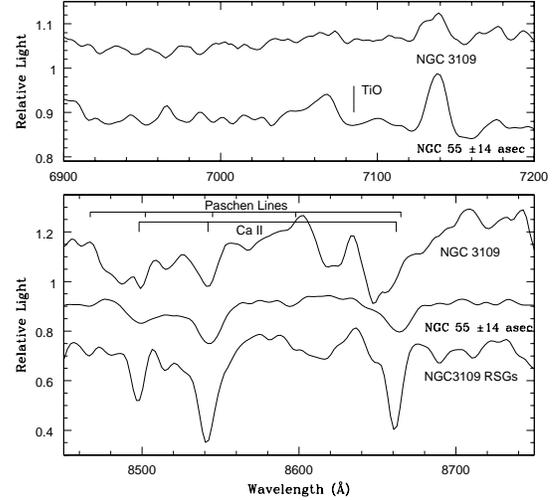}
\caption{Mean spectra of NGC 3109, the central regions 
of NGC 55, and of the three RSGs detected in NGC 3109. 
The wavelength interval in the top panel was selected for analysis as it is 
free of telluric emission signatures, while the lower panel examines 
wavelengths near the Ca triplet, where residuals from 
the subtraction of sky emission contributes noise 
to the NGC 3109 spectrum. The TiO band head near 7100\AA\ is weaker in NGC 3109 
than in the NGC 55 spectrum, while the Ca 8540\AA\ line in NGC 3109 -- which is 
not affected by strong telluric emission lines -- has a depth that 
is similar to that in NGC 55.}
\end{figure}

	The TiO band head near 7100\AA\ is weak in the NGC 3109 spectrum when 
compared with that seen in the NGC 55 spectrum. This indicates that evolved M stars 
contribute less light to the NGC 3109 spectrum near 7100\AA\ than in NGC 55, 
signalling a difference in stellar content. Residuals from the subtraction of telluric 
emission lines are significant throughout much of the wavelength region shown 
in the lower panel of Figure 19. While the Ca lines at 8498 and 8662\AA\ are 
distorted by telluric emission residuals, the Ca line at 8540\AA\ is 
in a part of the spectrum that is free of sky emission lines. 
Despite having a lower metallicity than NGC 55, Ca 8540 
is deeper in NGC 3109 than in the NGC 55 spectrum.

	The NGC 3109 spectrum is compared with SSP EMILES models in Figure 20. The 
models assume [M/H] = --0.66 for consistency with the metallicity measured 
among bright stars in NGC 3109 by Hosek et al. (2014). The TiO bandhead at 7100\AA\ 
becomes less pronounced in the models as age decreases, suggesting 
that the lack of an obvious TiO feature in NGC 3109 is an age effect. 
The depth of Ca 8540 is best matched by the 30 Myr model in Figure 20, which 
is the youngest age in the EMILES compilation. While based on 
only two features, these results are consistent with 
the notion that the red light in NGC 3109 is dominated by stars with ages $< 30$ Myr, 
as expected if there has been a recent large upswing in the star formation 
rate of NGC 3109. Dolphin et al. (2005) found evidence for a surge in recent 
star formation in the analysis of the NGC 3109 CMD.

\begin{figure}
\figurenum{20}
\epsscale{1.0}
\plotone{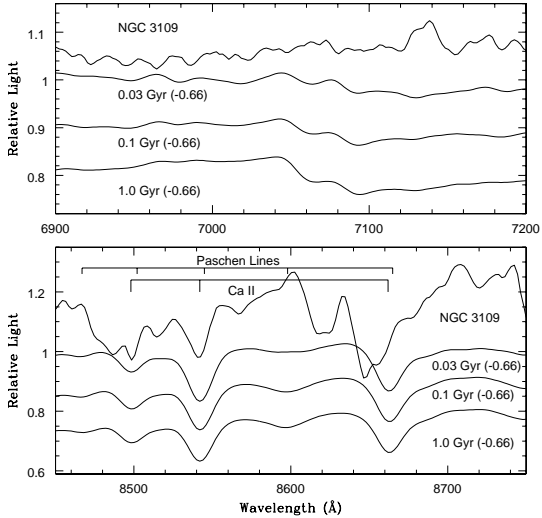}
\caption{EMILES model spectra with [M/H] $= -0.66$ are compared with the 
NGC 3109 spectrum in wavelength intervals that sample the TiO 
band head (top panel) and the Ca triplet (bottom panel). The 
TiO band head in the models weakens as age decreases. 
The Ca 8540 line is deeper than in the 0.1 and 1 Gyr models. 
With the caveat that the interpretation is based on only two spectral features, 
models with [M/H] $=-0.66$ and and age $< 0.1$ Gyr thus appear to better match the 
NGC 3109 spectrum than older models.}
\end{figure}

\section{DISCUSSION}

	The spectroscopic properties of the intermediate age LMC 
cluster NGC 1978 and the galaxies NGC 55 and NGC 3109 have been examined in 
the wavelength interval $0.7 - 1.1\mu$m using long slit spectra 
recorded with GMOS on the Gemini South telescope. Spectra of integrated light from 
each system have been extracted, and these have been compared with 
model spectra from the EMILES compilation. Spectra of individual bright stars have 
also been extracted from the NGC 1978 and NGC 3109 spectra -- NGC 55 is 
too distant for light from individual stars to be extracted from these data. In 
the case of NGC 1978 the stellar spectra allow the influence that individual bright 
stars have on the integrated spectrum to be explored.

	NGC 1978 is an intermediate age ($\sim 2$ Gyr) cluster with a 
metallicity that is comparable to that of NGC 55. There has been debate 
during the past decade about the nature of intermediate age clusters 
in the Magellanic Clouds, with emphasis on the origin of 
the extended main sequence turn-offs that have been found in the CMDs of many of these 
(see Goudfrooij et al. 2017 \& Li et al. 2016 for recent discussions). 
Published spectroscopic and photometric observations indicate that 
NGC 1978 does not have an extended main sequence turn-off and that there is a 
tight upper limit on star-to-star metallicity differences. These argue that 
NGC 1978 is a SSP. 

	The NGC 1978 observations consist of five contiguous slit pointings, made 
through 2 arcsec wide slits that are aligned parallel to the major axis. 
Raster-scanning -- in which a target is moved across 
the slit during an observation -- is another technique that can be used to obtain the 
integrated spectra of star clusters. However, while raster-scanning has the potential 
of sampling light from a large part of a cluster, the spatial information that is 
needed to examine contributions from individual bright stars is lost.

	The CaT indices measured in the spectra of 13 evolved stars 
in NGC 1978 are consistently larger than is expected for K and M giants 
with the same metallicity as NGC 1978. Deeper than expected Ca 
lines could occur if stars in NGC 1978 have [Ca/Fe] 
$> 0$. However, observations challenge such an explanation for NGC 1978. 
Olszewski et al. (1991) discuss Ca measurements of NGC 1978 stars 8 and 9 in 
the numbering scheme of Lloyd Evans (1980b). Figure 10 of Olszewski et al. (1991) 
demonstrates that these stars do not depart from the trend between brightness and 
Ca equivalent width defined by other stars in the Magellanic Clouds.  
Star 8 is 2MASS05284847--6614387, which was one of the stars extracted from the 
GMOS spectra. Aside from the likely foreground stars 2MASS05283756--6613040 and 
2MASS05284668--6614262, and the C star 2MASS05284449--6614039, 2MASS05284847--6614387 
has the second lowest CaT index listed in Table 2. Both of the stars observed 
by Olszewski et al. (1991) in NGC 1978 are in the outer regions of the 
cluster (Plate 1 of Lloyd Evans 1980b), raising the possibility 
that one or both may belong to the LMC field population. 
[Ca/Fe] $\sim 0$ among intermediate age stars in the 
LMC bar and inner disk (Van der Swaelmen et al. 2013). 

	Mucciarelli et al. (2008) examine the chemical compositions 
of 11 bright giants in NGC 1978, and find a mean [Ca/Fe] $= -0.11$ with a dispersion 
$\pm 0.05$ dex. They also measure abundances for the $\alpha$ elements 
O, Mg, and Si. The average [Mg/Fe] and [Si/Fe] is above solar, while 
[O/Fe] is sub-solar. As with the other intermediate age clusters examined by 
Mucciarelli et al. (2008) it thus appears that [$\alpha$/Fe] in NGC 1978 is close to 
solar. Thus, these measurements do not support a high [Ca/Fe] in NGC 1978. 

	The deep Ca lines in the NGC 1978 spectrum are not the only 
peculiar properties associated with this cluster. Lederer et al. (2009) discuss 
the abundances of C isotopes and O in bright stars in NGC 1978, and find differences 
when compared with models and similar measurements in the intermediate age cluster 
NGC 1846. They consider three possible causes for the abundance peculiarities 
in NGC 1978: multiple populations, stellar rejuventation, and extra mixing. As noted 
above, there are indications that NGC 1978 is an SSP, arguing against the first and 
second of these causes. As for extra mixing, Lederer at al. (2009) emphasize the 
{\it ad hoc} nature of this explanation, and highlight that much work needs to be 
done to better understand mixing mechanisms. Nevertheless, we note that extra mixing 
could be a consequence of rotation. If stars in NGC 1978 had a uniformly high 
initial rotation rate that persisted for much of their lives then they will 
develop larger core masses than non-rotating stars. This in turn could drive the 
surface gravities during the advanced stages of evolution to lower values than 
in non-rotating stars. It should be emphasized that there is no direct 
evidence for a uniformly high initial spin rate among stars in NGC 1978, and so this 
explanation is speculative only. Still, the prospect that the comparatively 
strong Ca triplet lines in the integrated spectrum of NGC 1978 might be linked to 
the chemical peculiarities found by Lederer et al. (2009) warrants further exploration.

	While the depths of Ca triplet lines in model spectra do not match those 
observed in NGC 1978, the agreement between models and the spectra of NGC 55, which 
has a metallicity that is similar to NGC 1978, is better. 
NGC 55 is of course not an SSP, as stars in NGC 55 formed over a range 
of epochs. Still, comparisons with SSP models are useful to establish a characteristic 
age of the stars that contribute most to the light at these wavelengths. 
Comparisons with the models suggest that a large fraction of the integrated light at 
red wavelengths comes from populations with ages 1 -- 2 Gyr. The CN bandhead 
near 7900\AA\ is present but weak in the NGC 55 spectrum, suggesting that C stars 
do not contribute significantly to the integrated 
light in the area sampled by these spectra. Davidge (2005) 
discussed the star-forming history of NGC 55, and used different lines of evidence to 
conclude that the star formation rate in NGC 55 during recent epochs is much lower 
than during intermediate epochs. The characteristic age deduced from the 
GMOS spectra is consistent with this.

	The GMOS spectra cover $\pm 44$ arcsec ($\sim 0.5 - 0.6$ kpc) on either side 
of the major axis of NGC 55, and a gradient in the depth of the NIR Ca 
triplet is seen. Davidge (2005) found a gradient in the photometric properties of the 
most luminous AGB stars, albeit over a considerably larger range of offsets 
from the disk plane than are examined with the GMOS spectra. 
The disk of NGC 55 is inclined by 79$^o$ to the line of sight and the 
Holmberg radius is $\sim 20$ arcmin (Puche et al. 1991). The GMOS spectra thus 
sample light from the disk, even at the largest offsets, and the variation 
in CaT is likely driven by radial metallicity gradients in the NGC 55 disk. 
Indeed, Kudritzki et al. (2016) find a radial metallicity gradient among young blue 
supergiants in the NGC 55 disk. The GMOS data suggest that a metallicity gradient also 
is present among the intermediate age stars that dominate the integrated spectrum.

	Non-secular processes may have played 
a significant role in shaping the present-day properties of the NGC 55 disk. 
Kudritzki et al. (2016) examine the chemical evolution of the NGC 55 disk, 
and find evidence for large scale infall and outflow. 
Tanaka et al. (2011) find metal-poor structures in the outermost 
regions of NGC 55, and suggest that these are artifacts of mergers with dwarf 
satellites. Tidal interactions during the assimilation of such 
companions may heat the disk. There is also evidence for interactions 
during intermediate epochs that could have triggered a large star-forming event. 
Westmeier et al. (2013) find an asymmetric HI distribution associated with NGC 55 
that is interpreted as an artifact of ram pressure interactions with the inter-galactic 
medium, with the HI compacted in the western part of the galaxy, as expected if NGC 55 
were moving away from NGC 300, which is its nearest large neighbor. They also find 
evidence that the gaseous component of the NGC 55 disk has been stirred. These two 
findings lead Westmeier et al. (2013) to suggest that NGC 55 and NGC 300 interacted 
1 -- 2 Gyr in the past, at which time the disk of NGC 55 was tidally heated. An 
interaction of this nature might also induce elevated levels of star-forming 
activity, and the timing of the interaction between the two galaxies is 
consistent with the characteristic age of the red spectrum of NGC 55. 

	The GMOS slit passes through one of the two star-forming complexes 
identified by Ferguson et al. (1996). While emission from H$\alpha$ and other 
lines is widespread near the major axis of NGC 55, 
the spectrum of the part of cluster 1 that is sampled by GMOS contains 
line emission that is much weaker than in its immediate surroundings. 
Thus, this part of cluster 1 likely does not harbor a 
large concentration of massive hot ionizing stars. The presence of deep Ca 
absorption features indicates an age of at least 8 Myr for cluster 1, 
with an estimated upper age limit of 30 Myr. 
A deep CN bandhead at 7900\AA\ is also seen in the cluster 1 
spectrum. While this may be the result of an interloping C star (or interloping 
C stars), a moderately strong CN bandhead is also present in model spectra with ages 
of a few tens of Myr (e.g. Figure 16), suggesting that it may originate in RSGs. 
If C stars are present then C$_2$ absorption should be seen at $1.85\mu$m. 
Davidge (2005) found a population of bright red 
supergiants in cluster 1, and the location of these objects on 
the $(J-H, H-K)$ two color diagram is consistent with an age of at least 10 Myr. 
Thus, there is consistency between the photometric and spectroscopic age estimates of 
cluster 1. 

	We close the discussion by examining NGC 3109. 
Holtzman et al. (2005) discuss a moderately deep CMD 
of a field in NGC 3109 that is restricted to stars 
with M$_V < 0$. The morphology of the CMD is consistent with a constant SFR over 
a range of epochs, followed by a factor of 2 upswing in the SFR during the 
past 0.1 Gyr (Dolphin et al. 2005). While the cause of a recent upturn in 
star-forming activity in NGC 3109 is a matter of speculation, Barnes \& de Blok 
(2001) suggest that NGC 3109 and Antlia A may have interacted $\sim 1$ Gyr ago. 
An interaction between these galaxies could have sparked elevated levels of 
star-forming activity at a time that coincides with the rise in the SFR indicated by 
the CMDs.

	The interpretation of the NGC 3109 spectra is 
complicated by the low S/N ratio of the data, and the age information 
found from these spectra should be considered 
as preliminary only. Still, while there is the 
potential for an age-metallicity degeneracy, if [Fe/H] $= -0.66$ is adopted 
for the models as indicated by the abundance analysis of some of the brightest members 
of the galaxy (Hosek et al. 2014) then the models point to a characteristic age 
of tens of Myr, as opposed to the 1 -- 2 Gyr age 
found for NGC 55. This conclusion is based on the depth of the Ca8540 
line, which is sharper and deeper than its counterpart in the NGC 55 spectrum, 
despite originating in a lower metallicity environment. This 
suggests a larger contribution to the light from RSGs in NGC 3109 than in NGC 55. 
The weak TiO bandhead near 7100\AA\ also supports 
a characteristic age $< 0.1$ Gyr. That significant star-forming activity 
has occured during recent epochs is consistent with the conclusion reached by 
Dolphin et al. (2005) that the SFR in NGC 3109 experienced an upturn 
during the past 0.1 Gyr. 

\section{CONCLUSIONS}

	Based on the spectra presented in this paper, we conclude the following:

\parindent=0.0cm

1) Pointing-to-pointing differences are seen in the 
NGC 1978 spectra, highlighting the role that stochastic effects play at red 
and NIR wavelengths. The most obvious differences in the overall appearance 
of the spectra are attributed to the contribution made by the C star 
2MASS05284449--6614039 to the light in each pointing. These differences 
highlight that the CN band head at 7900\AA\ can signal the presence of C stars 
in the integrated spectra of intermediate age systems.

2) Spatial variations in the distribution of the brightest M giants can 
significantly affect the strengths of features such as the Ca triplet in slit 
spectra that sample only a small percentage of the light from the cluster. 
The impact of these variations is of course muted when light from different 
pointings is combined. Indeed, the CaT index defined by Cenarro et al. (2001a) is 
not greatly affected by the presence of the C star or the brightest M giant in 
the combined spectrum from all 5 pointings, changing by only 
0.1\AA\ when either star is removed. This demonstrates that the CaT index is 
a robust probe of red stellar content in dense stellar systems like NGC 1978.

3) The CaT indices measured in the spectra of the most evolved stars in NGC 1978 are 
consistently larger than is expected based on Galactic K and M giants that have the 
same metallicity as NGC 1978. Model spectra from the EMILES compilation that adopt the 
age and metallicity of NGC 1978 also underestimate the depth of the Ca triplet lines. 
Evidence is presented that this is not due to [Ca/Fe], and it 
is suggested that rotation may have affected the surface gravities of the most 
evolved stars in NGC 1978.

4) Surface photometry measurements made from 2MASS 
images reveal a central red cusp within $\pm 10$ arcsec 
of the center of NGC 1978, which is a region that contains some of the brightest stars 
in the cluster. While 2MASS05284449--6614039 contributes a significant fraction of 
the light from this part of the cluster, this cusp is not due solely to that star. 
Rather, the red cusp is a probable signature of mass segregation in 
NGC 1978, which might be primordial in origin (Parker et al. 2016). 
The stronger than expected Ca line strengths are not due to mass segregation, as 
the depth of the Ca triplet does not change if a spectrum is 
constructed that excludes light from the central cusp.

5) In contrast with NGC 1978, model spectra reproduce the depths 
of the Ca lines in the spectrum of NGC 55. Comparisons with the 
models suggest that a large fraction of the integrated light from NGC 55 at 
red wavelengths comes from populations with ages 1 -- 
2 Gyr. The CN bandhead near 7900\AA\ is present but weak in the NGC 55 spectrum, 
suggesting that C stars do not contribute significantly to the integrated 
light in the area sampled by these data.

6) A gradient in the depth of the NIR Ca triplet with respect to distance 
from the major axis is seen in the NGC 55 spectrum. This 
is likely due to a radial metallicity gradient in the NGC 55 disk.

7) The spectrum of the part of cluster 1 in NGC 55 that is sampled by GMOS contains 
line emission that is much weaker than in its immediate surroundings. 
The presence of deep Ca absorption features indicates an age of at least 8 Myr 
for cluster 1, with an estimated upper age limit of 30 Myr. 

8) Comparisons with models suggest a characteristic age 
of tens of Myr for NGC 3109, as opposed to the 1 -- 2 Gyr age 
found for NGC 55. This age estimate is based largely on the depth of the Ca8540 
line, which is sharper and deeper than its counterpart in the NGC 55 spectrum.

\acknowledgements{Thanks are extended to the anonymous referee for providing a 
report that improved the content and presentation of the paper.}

\end{document}